\documentclass[aps, prb, twocolumn, superscriptaddress,preprintnumbers,10pt]{revtex4-2}
\usepackage{bm, amsmath, amsfonts, amssymb, ascmac, mathtools, braket}
\usepackage{multirow}
\usepackage{graphicx}
\usepackage{float, color, xcolor}

\definecolor{rred}{rgb}{0.8, 0.0, 0.0}
\definecolor{bblue}{rgb}{0.0, 0.0, 0.8}

\usepackage[whole]{bxcjkjatype} 
\usepackage{subfigure} 
\usepackage{amscd} 

\usepackage{bbm}
\usepackage{tabularx}

\usepackage{comment}

\usepackage[
pagebackref=false,
colorlinks=true,
linkcolor=bblue,
urlcolor=bblue,
filecolor=black,
citecolor=rred,
pdfstartview=FitV,
pdftitle={},
pdfauthor={},
pdfsubject={},
pdfkeywords={},
pdfpagemode=None,
bookmarksopen=true
]{hyperref}

\begin{document}

\title{Complex nonlinear sigma model}

\author{Kazuki Yamamoto}
\email{kazuki-yamamoto@omu.ac.jp}
\affiliation{Research Institute for Innovation and Co-Creation, Osaka Metropolitan University, Sakai, Osaka 599-8531, Japan}
\affiliation{Department of Physics, Osaka Metropolitan University, Sumiyoshi, Osaka 558-8585, Japan}
\affiliation{Nambu Yoichiro Institute of Theoretical and Experimental Physics (NITEP), Osaka Metropolitan University, Sumiyoshi, Osaka 558-8585, Japan}
\affiliation{Department of Physics, Institute of Science Tokyo, Meguro, Tokyo 152-8551, Japan}

\author{Kohei Kawabata}
\email{kawabata@issp.u-tokyo.ac.jp}
\affiliation{Institute for Solid State Physics, University of Tokyo, Kashiwa, Chiba 277-8581, Japan}

\date{\today}

\begin{abstract}
Motivated by the recent interest in the criticality of open quantum many-body systems, we study nonlinear sigma models with complexified couplings as a general framework for nonunitary field theory.
Applying the perturbative renormalization-group analysis to the tenfold symmetric spaces, we demonstrate that fixed points with complex scaling dimensions and critical exponents arise generically,  
without counterparts in conventional nonlinear sigma models with real couplings. 
We further clarify the global phase diagrams in the complex-coupling plane and identify both continuous and discontinuous phase transitions. 
Our work thus identifies nonlinear sigma models as a representative setting for studying complex critical points and elucidates universal aspects of critical phenomena in complexified field theory.
\end{abstract}

\maketitle

\section{Introduction}

Phase transitions and critical phenomena provide clear perspectives on universality~\cite{Goldenfeld-textbook, Cardy-textbook, Sachdev-textbook}.
Near criticality, microscopic details become irrelevant, and systems as diverse as classical fluids, quantum magnets, and gauge theories exhibit identical long-distance behavior governed by a small set of universal data. 
A central theoretical framework for understanding this emergent universality is renormalization, which systematically tracks how coupling constants flow under coarse graining and reveals how fixed points encode scaling laws and critical exponents. 
In practice, much progress has been driven by detailed studies of paradigmatic models. 
For example, the Ising model is the canonical example of a theory with $\mathbb{Z}_2$ discrete symmetry that can be spontaneously broken.

For continuous symmetry, the nonlinear sigma model plays an analogous role: 
\begin{equation}
    S = \frac{1}{2t} \int d^d x\,\mathrm{tr} \left[ \left( \nabla Q^{\dag} \right) \left( \nabla Q \right) \right],
        \label{eq: NLSM}
\end{equation}
where an $N \times N$ matrix field $Q$ is constrained to lie on a target manifold, and the action is built from the simplest symmetry-allowed gradient term with the coupling constant $t$. 
The target spaces are organized into the tenfold symmetry classes~\cite{Zirnbauer-96, *AZ-97}, further determining the distinct universality classes.
Historically, nonlinear sigma models were developed as indispensable tools in high energy physics, capturing the low-energy dynamics of Nambu-Goldstone modes and furnishing controlled settings for renormalization and asymptotic behavior~\cite{Polyakov-75, Brezin-76L, *Brezin-76B}. 

In condensed matter physics, nonlinear sigma models provide the effective field theory~\cite{Wegner-79, *Schafer-80, Gorkov-79, Hikami-80, Efetov-80, Altshuler-80} of Anderson transitions in disordered electron systems~\cite{Dyson-53, Anderson-58, Abrahams-79, Lee-review, Evers-review},
where the classification of possible topological terms corresponds to the classification of topological insulators and superconductors~\cite{Schnyder-08, Kitaev-09, Ryu-10, CTSR-review}.
They also underlie the low-energy description of quantum antiferromagnets and play a significant role in the Haldane gap phenomena~\cite{Haldane-83PLA, *Haldane-83PRL, Affleck-89, Tasaki-textbook}. 
More recently, nonlinear sigma models have reappeared in effective descriptions of measurement-induced phase transitions of free fermions~\cite{Jian-22, *Jian-23, Yang-23, Fava-23, Poboiko-23}, 
further reinforcing their status as a prototypical framework for critical phenomena.

Until recently, the theory of criticality was formulated largely for isolated equilibrium systems described by Hermitian Hamiltonians. 
In parallel with the rapid development of nonequilibrium many-body physics, however, there has been growing interest in phase transitions and critical phenomena in open systems effectively governed by non-Hermitian generators, where interactions with external environments make coupling constants intrinsically complex valued~\cite{Konotop-review, Christodoulides-review}. 
A paradigmatic manifestation is the spontaneous breaking of parity-time symmetry accompanied by a real-to-complex spectral transition~\cite{Bender-02, Bender-review}. 
This traces back to the Yang-Lee theory, which studies zeros of the partition function for the Ising model with an imaginary magnetic field~\cite{Yang-52, *Lee-52, Fisher-78, Cardy-23}.
In $1+1$ dimensions, the associated critical behavior is described by nonunitary conformal field theory with a negative central charge~\cite{Cardy-85, Bianchini-15, Couvreur-17, Chang-20, Lee-22, Tu-22, Ryu-23, Hsieh-23, Fossati-23, Rottoli-24, Xue-25, Li-25, Fan-25, Cruz-25, Miro-25, Chou-25}.

Furthermore, analytic continuation of coupling constants into the complex plane yields nonunitary critical theory characterized by genuinely complex universal data~\cite{Kaplan-09, Wang-17, Gorbenko-18a, *Gorbenko-18b, Benini-20, Faedo-20, Giombi-20, Gorbenko-20, Nahum-22, Han-23, Haldar-23, Yang-26}, 
with roots tracing back to the study of Fisher zeros~\cite{Fisher-65}.
Concrete lattice realizations have recently begun to illuminate this regime.
For example, the weakly first-order phase transition of the five-state quantum Potts model in $1+1$ dimensions can, upon complexifying parameters, be related to a pair of complex fixed points and an underlying complex conformal field theory with a complex central charge~\cite{Ma-19, Jacobsen-24, Tang-24, Shimizu-25, Linden-25, Tang-25}. 
Yet, despite these advances, complexified field theory remains understood only through a relatively small set of examples.
A more systematic understanding therefore requires elucidating how renormalization-group flows, fixed points, and critical scaling reorganize in nonunitary field theory. 

In this work, we develop a perturbative renormalization-group description of nonlinear sigma models with complexified coupling constants across the tenfold symmetry classes of target manifolds. 
Building on established beta functions, we analytically continue the coupling into the complex plane and systematically investigate the resulting flows, fixed points, and scaling dimensions in one, two, and three dimensions. 
We show that complex fixed points emerge generically as complex-conjugate pairs 
and that the associated universal data, such as scaling dimensions and critical exponents, become genuinely complex. 
This complex universality is directly reflected in the spiral structure of the renormalization-group flows, with no analogs in conventional nonlinear sigma models with real couplings.
Going beyond the local analysis near fixed points, we further elucidate the global phase structures in the complex-coupling plane and identify both continuous and discontinuous phase transitions. 

Overall, our results provide characterization of phase structure and criticality in field theory with complexified couplings, establishing complex nonlinear sigma models as a broadly applicable framework for nonunitary critical phenomena in open quantum many-body systems. 
Our results therefore extend the scope of complexified field theory beyond the few examples previously studied, such as the non-Hermitian Potts model, and provide useful guidance for constructing and interpreting microscopic non-Hermitian lattice models that realize complex criticality.
While our analysis is based on the finite-order perturbative renormalization group, several qualitative features are robust, 
such as the generic emergence of complex-conjugate fixed points and the associated spiral renormalization-group flows governed by complex scaling dimensions.

The rest of this work is organized as follows. 
In Sec.~\ref{sec: On}, we study in detail the perturbative renormalization group of the $\mathrm{O} \left( N \right)$ nonlinear sigma model with complex couplings as an illustrative case.
We characterize the complex fixed points and scaling dimensions induced by complexification, and reveal the global phase structures in one, two, and three dimensions.
In Sec.~\ref{sec: other}, we systematically extend this analysis to the remaining nine symmetric spaces of the complex nonlinear sigma models.
Finally, in Sec.~\ref{sec: conclusion}, we conclude this work with several outlooks.

\section{\texorpdfstring{$\mathrm{O} \left( N \right)$}{On}}
    \label{sec: On}

As a representative case, we begin with studying the nonlinear sigma model in Eq.~\eqref{eq: NLSM} with target manifold given by the orthogonal group $\mathrm{O} \left( N \right)$.
Using the perturbative renormalization-group equation, we demonstrate that the complex-coupling plane generically supports fixed points with complex scaling dimensions.
We further elucidate the global phase diagrams in one, two, and three dimensions.

\subsection{Perturbative renormalization group}

On the basis of the perturbative theory carried out up to the fifth order in the coupling constant $t$, the renormalization-group equation for the $\mathrm{O} \left( N \right)$ nonlinear sigma model is given by~\cite{Hikami-81, Wegner-89}
\begin{align}
\frac{dt}{dl} &= \beta \left( t \right) = \left( 2-d \right) t + \left( N-2 \right) t^2 + \frac{1}{2} \left( N-2 \right)^2 t^3 \nonumber \\
&\qquad +\frac{3}{8} \left( N-2 \right)^3 t^4 - \left( N-2 \right) c_1 \left( N \right) t^5,
    \label{eq_On}
\end{align}
with
\begin{align}
    c_1 \left( N \right) &\coloneqq - \left[ \left( \frac{19}{48} + a \right) \left( N - 2 \right)^3 \right. \nonumber \\ 
    &\qquad\qquad - a \left( N - 3 \right) \left( N - 4 \right) \left( N + 2 \right) \bigg], \label{eq_coefficient_c1}\\ 
    a &\coloneqq \frac{3\zeta \left( 3 \right)}{16} = 0.225386 \cdots. \label{eq_coefficient_a}
\end{align}
Additionally, $l \coloneqq \log L$ denotes the logarithm of the system length $L$.
While Eq.~\eqref{eq_On} is usually considered for a real coupling $t \in \mathbb{R}$, we here analytically continue it to a complex coupling $t \in \mathbb{C}$.
Such a complexified theory should arise as an effective field theory of microscopic lattice models described by non-Hermitian Hamiltonians.
For example, a recent study has shown that a complex $\mathrm{O} \left( 3 \right)$ nonlinear sigma model emerges as an effective description of a non-Hermitian quantum antiferromagnetic chain~\cite{Yang-26}.

\begin{figure*}[t]
\includegraphics[width=1.0\linewidth]{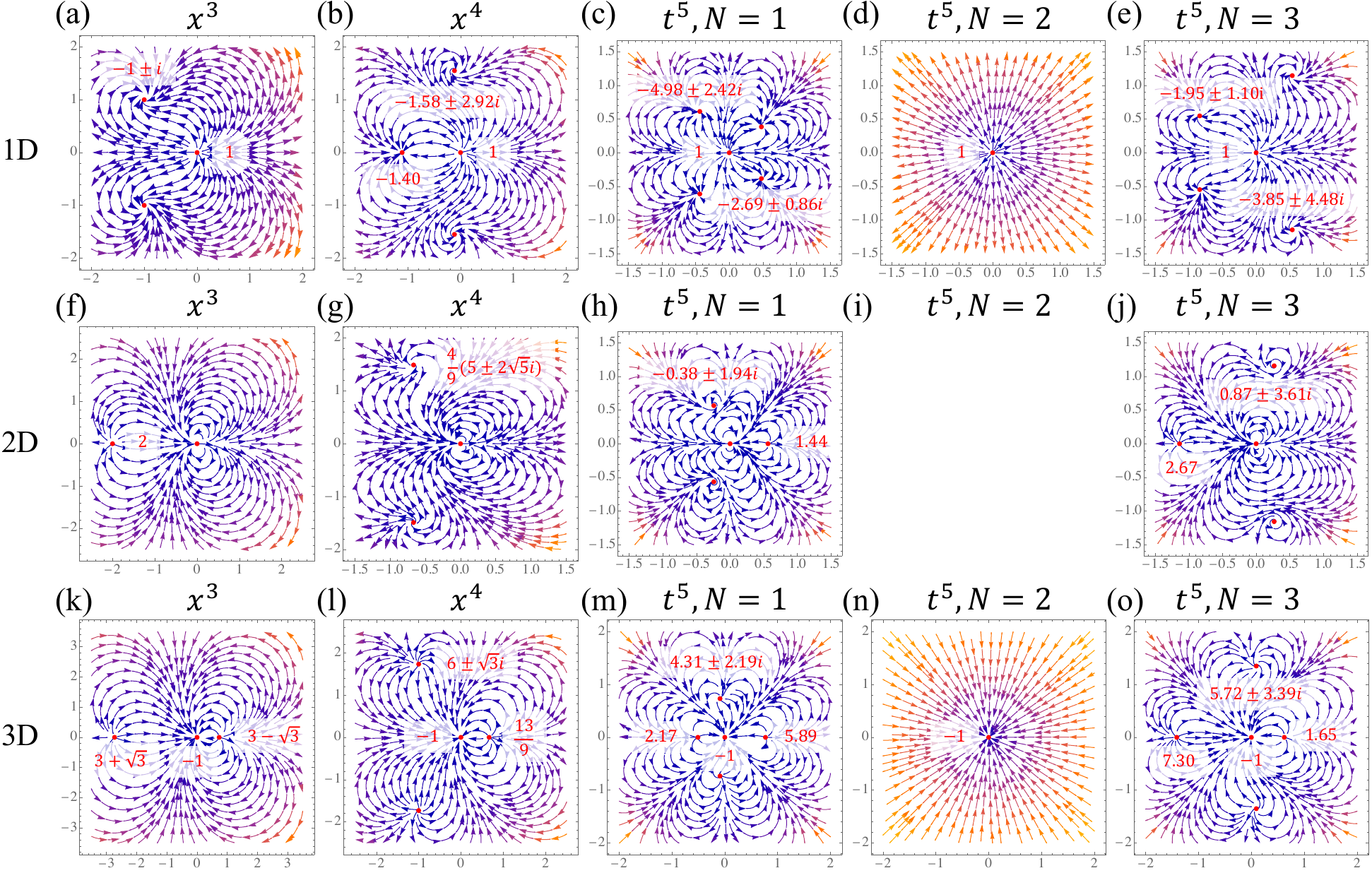}
\caption{Perturbative renormalization-group flow of the $\mathrm{O} \left( N \right)$ nonlinear sigma model in the complex-coupling plane. 
Fixed points (red points) occur either on the real axis or in complex-conjugate pairs, and their scaling dimensions are also shown.
The top, middle, and bottom panels correspond to one, two, and three dimensions, respectively. 
The first and second columns show the results truncated at the cubic and quartic order in perturbation theory for the rescaled coupling $x \coloneqq \left( N-2 \right) t$. 
The third, fourth, and fifth columns present the quintic-order results for $N=1$, $N=2$, and $N=3$, respectively. 
In one dimension, all complex fixed points are stable. 
In two and three dimensions, unstable complex fixed points appear beyond the quartic order.}
    \label{fig_On}
\end{figure*}

We then analyze the resulting renormalization-group flow in the complex-coupling plane, as shown in Fig.~\ref{fig_On}.
We present the results in one [Figs.~\ref{fig_On}\,(a)-(e)], two [Figs.~\ref{fig_On}\,(f)-(j)], and three [Figs.~\ref{fig_On}\,(k)-(o)] dimensions. 
We also compare the different truncations of perturbation theory in Eq.~\eqref{eq_On}: 
the first column retains the terms up to $\mathcal{O}\,( t^3 )$, 
the second up to $\mathcal{O}\,( t^4 )$, 
and the third, fourth, and fifth include the terms up to $\mathcal{O}\,( t^5 )$ for $N=1$, $N=2$, and $N=3$, respectively.
Although Eq.~\eqref{eq_On} depends explicitly on $N$, it can be cast into an $N$-independent form, up to the fourth order $\mathcal{O}\,( t^4 )$, by rescaling the coupling $t$ as $x \coloneqq \left( N-2 \right) t$.
Accordingly, the renormalization-group flow does not depend on $N$ for $\mathcal{O}\,( t^4 )$.
For $N=2$ and $d=1, 3$, the beta function $\beta \left( t \right)$ vanishes except for the linear term $\left( 2-d \right) t$, yielding a purely radial flow away from or toward the origin [see Figs.~\ref{fig_On}\,(d), (n)].
For $d=2$, even the linear term vanishes, making the renormalization-group flow ill defined;
this is why no flow is shown in Fig.~\ref{fig_On}\,(i).

\subsection{Complex fixed points}

As shown in Fig.~\ref{fig_On}, fixed points emerge in the complex-coupling plane, which have no counterparts in the conventional nonlinear sigma model.
A distinctive feature of these complex fixed points is that they are associated with complex scaling dimensions.
As an illustrative example, let us consider the three-dimensional $\mathrm{O} \left( 3 \right)$ nonlinear sigma model within the fourth order of perturbation theory [see Fig.~\ref{fig_On}\,(l)]. 
The fixed points are given by the zeros of the beta function, $\beta \left( t \right) = 0$, i.e.,
\begin{equation}
    t_{\rm c} = 0, \frac{2}{3}, -1 \pm \sqrt{3}i.
\end{equation}
In addition to the real-coupling fixed points $t_{\rm c} = 0, 2/3$, a complex-conjugate pair of fixed points, $t_{\rm c} = -1\pm\sqrt{3}i$, appears upon complexifying the coupling.
To capture the renormalization-group flow in the vicinity of a complex fixed point $t_{\rm c} = -1 + \sqrt{3}i$, we linearize Eq.~\eqref{eq_On} as
\begin{equation}
    \frac{d\tilde{t}}{dl} \simeq y_t \tilde{t}, \quad y_t = \beta'\,( t_{\rm c} = -1+\sqrt{3}i ) = 6-\sqrt{3}i
\end{equation}
with $\tilde{t} \coloneqq t - \left( -1 + \sqrt{3}i \right)$.
It then follows that, near this fixed point, the complex coupling evolves by
\begin{equation}
    \tilde{t} \simeq \tilde{t}_0 e^{(6-\sqrt{3}i)\,(l-l_0)},
\end{equation}
given the initial condition $\tilde{t} \left( l=l_0 \right) = \tilde{t}_0$.
Accordingly, the complex coupling flows outward from this fixed point, rotating clockwise in the complex-coupling plane.
In the theory of dynamical systems, this behavior is referred to as an unstable spiral~\cite{Strogatz-textbook}.

In this manner, we calculate complex scaling dimensions for all the complex fixed points, shown in Fig.~\ref{fig_On}.
In general, the real part of the scaling dimension $y_t$ governs the stability of the fixed point:
it is unstable for $\mathrm{Re}\,y_t > 0$, whereas it is stable for $\mathrm{Re}\,y_t < 0$.
By contrast, the imaginary part of $y_t$ describes the angular velocity and rotation direction of the spiral renormalization-group flow.
Moreover, the complex scaling dimension $y_t$ is reflected in the critical behavior of the correlation length $\xi$.
Specifically, $\xi$ diverges around the critical point $t = t_{\rm c}$ as
\begin{equation}
    \xi \propto \left| t-t_{\rm c} \right|^{-\nu}, \quad \nu = \frac{1}{y_t}.
        \label{eq: nu}
\end{equation}
Owing to the complex-valued nature of $y_t$, 
the associated critical exponent $\nu$ is also complex.
While the real part of $\nu$ governs the power-law divergence $\propto \left| t-t_{\rm c}\right|^{-\mathrm{Re}\,\nu}$, the imaginary part causes the log-periodic oscillation, such as $\cos \left[ \left( \mathrm{Im}\,\nu \right) \log \left| t - t_{\rm c} \right| \right]$.
In general, $\left| t-t_{\rm c} \right|^{-\nu}$ in Eq.~\eqref{eq: nu} is complex.
This is consistent with the complex-valued nature of energy gaps in corresponding quantum systems, which scale as $\propto \xi^{-z} \propto \left| t - t_{\rm c}\right|^{z\nu}$ with the dynamical critical exponent $z$.

\subsection{Long-distance behavior}

The complex fixed points discussed so far dictate the renormalization-group flow around the origin.
In the long-distance (infrared) limit, it ultimately approaches either a stable fixed point or runs off to infinity in the complex-coupling plane.
To characterize this asymptotic behavior, we focus on the large-$\left| t \right|$ regime.
When the perturbation series is truncated at $\mathcal{O}\,( t^n )$ ($n \geq 2$), the renormalization-group equation asymptotically takes the form
\begin{equation}
    \frac{dt}{dl} \sim b_n t^n \quad \left( \left| t \right| \to \infty;\,b_n \in \mathbb{R} \right),
\end{equation}
which is solved as
\begin{equation}
    t \sim \frac{1}{\left[ t_0^{-(n-1)} - \left( n-1 \right) b_n \left( l-l_0 \right) \right]^{1/(n-1)}},
\end{equation}
under the initial condition $t \left( l=l_0 \right) = t_0$.
Although the $\left(n-1\right)$\,th root is generically multivalued with $n-1$ possible branches, the relevant branch is uniquely determined by the initial condition.
For generic initial conditions, the complex coupling $t$ flows toward the origin $t=0$ for $l \to \infty$.
For special initial conditions satisfying $t_0^{-(n-1)} b_n > 0$, however, it instead diverges to infinity in the complex plane.
Thus, there emerge $m-1$ asymptotic rays toward infinity:
\begin{equation}
    R_m: \begin{cases}
       re^{2m\pi i/(n-1)} & \left( b_n > 0\right); \\
       re^{(2m+1)\,\pi i/(n-1)} & \left( b_n < 0 \right), \\
    \end{cases}
        \label{eq: asymptote}
\end{equation}
for $r \gg 1$ and $m = 0, 1, 2, \cdots, n-2$.

These asymptotic rays generally arise in the complexified perturbative renormalization group in arbitrary dimensions.
In the vicinity of the origin, each ray is connected to an unstable fixed point, around which critical behavior arises.
Away from the fixed point, by contrast, the ray acts as a separatrix between distinct long-distance (infrared) fates: 
trajectories initialized on the ray flow to the strong-coupling regime $t\to\infty$, whereas an infinitesimal deviation drives the flow back to the weak-coupling basin $t\to 0$.
Accordingly, crossing the ray changes the long-distance endpoint of the flow discontinuously, which can be interpreted as a discontinuous phase transition.
This global structure is reminiscent of the phase diagram of the liquid-gas transition.
The asymptotes in Eq.~\eqref{eq: asymptote} follow directly from the perturbative renormalization-group equation, while it can break down for the strong-coupling regime.
It remains an open question whether a similar structure persists nonperturbatively.
In passing, the renormalization-group flow can, in principle, exhibit a limit cycle or even chaos.
However, we have not observed such exotic renormalization-group flows in our complexified nonlinear sigma models.

\subsection{One dimension}

Even in one dimension, we find fixed points unique to the complexified couplings [Figs.~\ref{fig_On}\,(a)-(e)].
However, all of the obtained complex fixed points are stable, implying the absence of the associated criticality.
For example, for $N=3$ at $\mathcal{O}\,( t^5 )$ [Fig.~\ref{fig_On}\,(e)], 
two complex-conjugate pairs of fixed points emerge, with the complex scaling dimensions $y_t = -1.95 \pm 1.10i$ and $y_t = -3.85 \pm 4.48i$.
Owing to $\mathrm{Re}\,y_t < 0$, these fixed points attract the nearby flows and form stable spirals.
In the real-coupling regime, the renormalization-group flow inevitably runs toward the strong-coupling regime with $t \to \pm \infty$.
In the complex-coupling regime, by contrast, generic trajectories converge toward the complex fixed points.
The physical implications of these emergent stable complex fixed points merit further investigation.

\subsection{Two dimensions}

In two dimensions, a complex-conjugate pair of fixed points generically appears [Figs.~\ref{fig_On}\,(f)-(j)], much as in one dimension.
However, in contrast to the one-dimensional case, these fixed points can have scaling dimensions with the positive real part and thus form unstable spirals.
For example, for $N=3$ at $\mathcal{O}\,( t^5 )$, the complex fixed points occur at $t_{\rm c} = 0.267 \pm 1.159i$ and are characterized by the complex scaling dimensions $y_t = 0.87 \pm 3.61i$.
Accordingly, these fixed points act as critical points unique to complexified field theory.
The asymptotic rays in Eq.~\eqref{eq: asymptote} that connect these complex fixed points to $\pm i \infty$ serve as lines of discontinuous phase transitions.
In Table~\ref{tab: 2D O(N) NLSM - tc}, we summarize the complex fixed points $t_{\rm c}$ for various values of $N$.
This order-by-order analysis shows that the qualitative structure is stable within the available perturbative orders.

\begin{table}[t]
	\centering
	\caption{Complex fixed points $t_{\rm c}$ for the two-dimensional complex $\mathrm{O} \left( N \right)$ nonlinear sigma model, obtained from the perturbative renormalization group at $\mathcal{O}\,( t^4 )$ and $\mathcal{O}\,( t^5 )$.}
	\label{tab: 2D O(N) NLSM - tc}
     \begin{tabular}{ccc} \hline \hline
    ~~~$N$~~~ & ~~~$t_{\rm c}$ [$\mathcal{O}\,( t^4 )$]~~~ & ~~~$t_{\rm c}$ [$\mathcal{O}\,( t^5 )$]~~~ \\ \hline
    $3$ & ~~~$-0.667 \pm 1.491i$~~~ & ~~~$0.267 \pm 1.159i$~~~ \\
    $4$ & $-0.333 \pm 0.745i$ & $0.134 \pm 0.579i$\\
    $5$ & $-0.222 \pm 0.497i$ & $0.082 \pm 0.414i$ \\
    $6$ & $-0.167 \pm 0.373i$ & $0.058 \pm 0.322i$ \\
    $7$ & $-0.133 \pm 0.298i$ & $0.045 \pm 0.262i$ \\
    $8$ & $-0.111 \pm 0.248i$ & $0.036 \pm 0.221i$ \\
    $9$ & $-0.095 \pm 0.213i$ & $0.031 \pm 0.190i$ \\
    $10$ & $-0.083 \pm 0.186i$ & $0.027 \pm 0.167i$ \\ \hline
    \end{tabular}
\end{table}

Complex conformal field theory behavior in the $\mathrm{O} \left( N \right)$ nonlinear sigma model has been analyzed in a recent work~\cite{Yang-26}.
In contrast to the perturbative approach employed in our work, it is based on the exact analysis of $\mathrm{O} \left( N \right)$ loop models using the Coulomb-gas method. 
Consistent with our perturbative results, it finds a complex-conjugate pair of critical points, whose critical exponents $\nu$ defined in Eq.~\eqref{eq: nu} are exactly obtained as
\begin{equation}
    \nu \left( N \right) = \frac{1}{4} \left( 1 \pm \frac{i\pi}{\mathrm{cosh}^{-1} \left( N/2 \right)} \right).
        \label{eq: CG}
\end{equation}
Although our perturbative renormalization-group analysis does not by itself establish the existence and precise properties of these complex critical points, 
it is consistent with this nonperturbative result and provides a complementary perspective.
Within our perturbative renormalization-group equation~\eqref{eq_On}, the same critical exponent $\nu$ at $\mathcal{O}\,( t^4 )$ is obtained as
\begin{equation}
    \nu = 0.25 \pm 0.224i,
        \label{eq: nu-t4}
\end{equation}
independently of $N$.
The real part of $\nu$ agrees with the exact result in Eq.~\eqref{eq: CG}, and the imaginary part is also of the same order of magnitude.
However, we also observe substantial deviations at $\mathcal{O}\,( t^5 )$, as summarized in Table~\ref{tab: 2D O(N) NLSM}.

We have also investigated whether Borel-Pad\'e resummation can stabilize the perturbative results.
See Appendix~\ref{asec: Borel-Pade} for details of the procedure, including the branch and contour prescriptions employed for the complex Borel integral and the choice of the Pad\'e approximants; 
see also, for example, Refs.~\cite{Ueoka-Slevin-14, *Ueoka-Slevin-17, TongWang-23}. 
For the Pad\'e approximant and analytic-continuation prescription adopted here, we find no nonzero complex zeros of the resummed beta function within the region examined.
This implies that the perturbative complex fixed points can be sensitive to this naive resummation procedure. 
However, only very low-order data are available in the present case, and the appropriate resummation scheme for complex fixed points in nonunitary field theory is itself nontrivial.
Accordingly, this result highlights a limitation of the present perturbative approach, and developing a fully controlled resummation method for nonunitary field theory remains an important open question.
As mentioned above, it should also be noted that the existence of complex fixed points is also supported by the nonperturbative analysis~\cite{Yang-26}.

\begin{table}[t]
	\centering
	\caption{Complex critical exponents $\nu$ for the two-dimensional complex $\mathrm{O} \left( N \right)$ nonlinear sigma model, obtained exactly by the Coulomb-gas method [Eq.~\eqref{eq: CG}]~\cite{Yang-26} and by the perturbative renormalization group at $\mathcal{O}\,( t^5 )$ [Eq.~\eqref{eq_On}].
    At $\mathcal{O}\,( t^4 )$, the critical exponent is obtained as $\nu = 0.25 \pm 0.224i$, independently of $N$ [Eq.~\eqref{eq: nu-t4}].}
	\label{tab: 2D O(N) NLSM}
     \begin{tabular}{ccc} \hline \hline
    ~~~$N$~~~ & ~~~Exact~~~ & ~~~Perturbative [$\mathcal{O}\,( t^5 )$]~~~\\ \hline
    $3$ & ~~~$0.25 \pm 0.816i$~~~ & ~~~$0.063 \pm 0.262i$~~~ \\
    $4$ & $0.25 \pm 0.596i$ & $0.063 \pm 0.262i$\\
    $5$ & $0.25 \pm 0.501i$ & $0.077 \pm 0.246i$ \\
    $6$ & $0.25 \pm 0.446i$ & $0.085 \pm 0.237i$ \\
    $7$ & $0.25 \pm 0.408i$ & $0.089 \pm 0.233i$ \\
    $8$ & $0.25 \pm 0.381i$ & $0.091 \pm 0.231i$ \\
    $9$ & $0.25 \pm 0.360i$ & $0.092 \pm 0.229i$ \\
    $10$ & $0.25 \pm 0.343i$ & $0.093 \pm 0.228i$ \\ \hline
    \end{tabular}
\end{table}

Additionally, Ref.~\cite{Yang-26} argues that the renormalization-group flow eventually runs to strong coupling.
This contrasts with our finite-order perturbative results, indicating that generic trajectories flow toward weak coupling $t=0$, while only special trajectories run to strong coupling $t \to \infty$.
Such a qualitative discrepancy may originate from genuinely nonperturbative effects that are invisible at any finite order in perturbation theory.  
In two dimensions, the beta function beyond perturbation theory can acquire contributions from nonperturbative saddles that are exponentially small at weak coupling yet can nevertheless reorganize the global flow.
These nonperturbative terms may change the basin structure of the complex renormalization-group flow and reconnect trajectories that appear to return to weak coupling within perturbation theory to runaway flows toward strong coupling, thereby reconciling the strong-coupling fate discussed in Ref.~\cite{Yang-26}.  
Clarifying this mechanism requires a nonperturbative analysis, for example, via large-$N$ methods.
We also note that the complex renormalization-group flow was studied on a finite two-dimensional lattice in the large-$N$ limit~\cite{Meurice-11}.

Another characteristic feature in two dimensions is the absence of the linear term, $\left( 2-d \right) t$, in the renormalization-group equation~\eqref{eq_On}, which underlies the asymptotic freedom in the real-coupling regime~\cite{Polyakov-75, Brezin-76L, *Brezin-76B}.
Notably, upon complexification, this generally leads to the circular renormalization-group flows:
a flow that initially departs from weak coupling $t=0$ eventually returns to it.
Indeed, the renormalization-group equation~\eqref{eq_On} around $t=0$, i.e., 
\begin{equation}   
    \frac{dt}{dl} = \left( N-2 \right) t^2,
        \label{eq: asymptotic freedom}
\end{equation}
generates an infinite family of circular flow lines passing through the origin, in addition to the flow line along the real axis.
By introducing the inversion $u \coloneqq 1/t$, Eq.~\eqref{eq: asymptotic freedom} reduces to $du/dl = - \left( N-2 \right)$ and yields a family of straight lines in the complex-$u$ plane with $\mathrm{Im}\,u = \mathrm{const}$.
Since inversion, as a M\"obius transformation, maps lines to a line or circles, Eq.~\eqref{eq: asymptotic freedom} correspondingly produces circular flows in the complex-$t$ plane.

Similar circular renormalization-group flows were also observed in the non-Hermitian Kondo models~\cite{Lourenco-18, Nakagawa-18, Gaiotto-21, Han-23}.
This is because the beta function has a closely analogous structure to that of the two-dimensional nonlinear sigma model.
Indeed, the perturbative renormalization-group equation for the Kondo model reads [see Eq.~(25) in Ref.~\cite{Nozieres-80}],
\begin{equation}
    \frac{d\Gamma}{d\log D} = - \rho \Gamma^2 + n\rho^2 \Gamma^3 + c\rho^3 \Gamma^4 + \cdots,
\end{equation}
which has the same form as Eq.~\eqref{eq_On} in two dimensions $d=2$.
This observation further underscores the ubiquity of the renormalization-group structures exhibited in Fig.~\ref{fig_On} for complexified field theory.

\subsection{Three dimensions}

In three dimensions, unstable fixed points appear even in the real-coupling regime [Figs.~\ref{fig_On}\,(k)-(o)].
For $N=3$ at $\mathcal{O}\,( t^5 )$ [Fig.~\ref{fig_On}\,(o)], the renormalization-group flow approaches weak coupling $t=0$ for $0 \leq t < t_{\rm c} = 0.622\cdots$ but strong coupling $t \to \infty$ for $t > t_{\rm c}$.
Upon the complexification, a complex-conjugate pair of fixed points newly appears, characterized by the complex scaling dimensions $y_t = 5.72 \pm 3.39i$.
As discussed above, generic trajectories flow toward the weak coupling $t = 0$.
In the vicinity of the complex fixed points, however, special trajectories can instead run to the imaginary strong coupling $t \to \pm i \infty$.
Crossing this curve, converging to the real or imaginary axis for $\left| t \right| \to \infty$, signals a discontinuous phase transition.

\section{Other symmetry classes}
    \label{sec: other}

Beyond the $\mathrm{O} \left( N \right)$ case, we investigate the complex nonlinear sigma model in Eq.~\eqref{eq: NLSM} with target manifolds given by the remaining nine symmetric spaces~\cite{Zirnbauer-96, *AZ-97}.
On the basis of the perturbative renormalization-group equations analytically continued to the complex plane in one, two, and three dimensions, we characterize the complex fixed points and scaling dimensions, as well as the global phase diagrams.

We also note that generic non-Hermitian matrices are classified into 38 symmetry classes~\cite{KSUS-19}, 
in contrast to the 10 symmetry classes for Hermitian matrices~\cite{Zirnbauer-96, *AZ-97}.
Nevertheless, this does not imply that the target manifolds of nonunitary nonlinear sigma models are enlarged from 10 to 38.
This is because the 38-fold classification of non-Hermitian matrices arises from the possible combinations of point-gap and line-gap classifying spaces~\cite{KSUS-19}. 
Each point-gap or line-gap classifying space itself is still one of the 10 symmetric spaces. 
Equivalently, from the perspective of spectral statistics, the 38-fold classification specifies the possible combinations of universal statistics for generic complex eigenvalues, eigenvalues near the real or imaginary axis, and eigenvalues near the origin; 
each of them is still governed by one of the 10 underlying symmetric spaces~\cite{Chen-25}.

\begin{figure*}[t]
\includegraphics[width=0.8\linewidth]{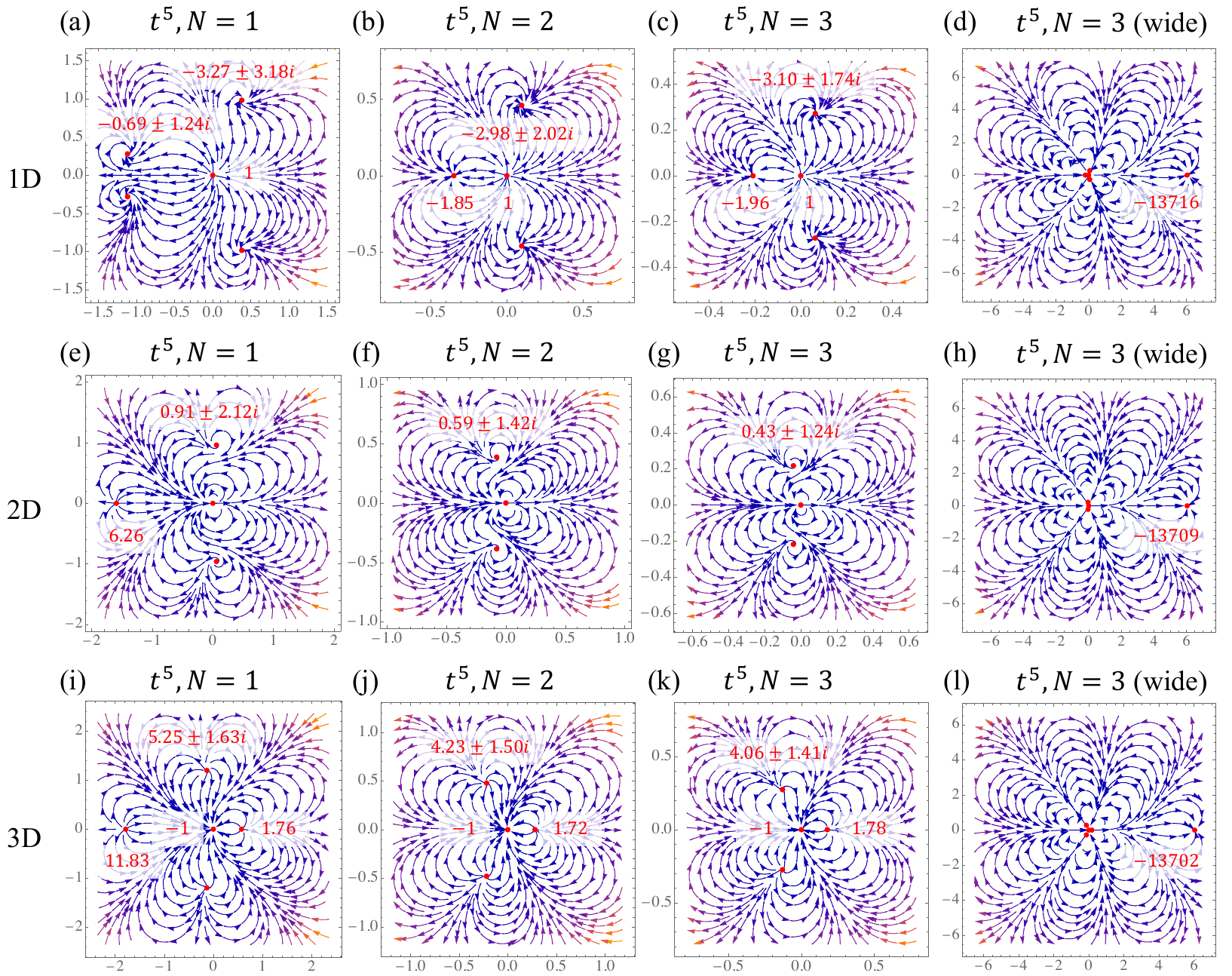}
\caption{Perturbative renormalization-group flow of the $\mathrm{U} \left( N \right)$ nonlinear sigma model in the complex-coupling plane at the quintic order in perturbation theory. 
Fixed points (red points) occur either on the real axis or in complex-conjugate pairs, and their scaling dimensions are also shown.
The top, middle, and bottom panels correspond to one, two, and three dimensions, respectively. 
The first and second columns show the results for $N=1$ and $N=2$, respectively. The third column presents the results for $N=3$ with a wide-range view shown in the fourth column.}
    \label{fig_Un}
\end{figure*}

\subsection{\texorpdfstring{$\mathrm{U} \left( N \right)$}{Un}}

For the nonlinear sigma model whose target space is the unitary group $\mathrm{U} \left( N \right)$, the renormalization-group equation up to the fifth order in perturbation theory is given by~\cite{Hikami-81, Wegner-89}
\begin{align}
\frac{dt}{dl} &= \left( 2-d \right) t + N t^2 + \frac{N^2}{2} t^3 +N^3\left[\frac{3}{8} +\left(\frac{19}{48}+a\right)N\right]t^4 \nonumber \\
&\qquad\qquad\qquad-aN^2 \left( N-2 \right)\left( N+2 \right) t^5,
    \label{eq_Un}
\end{align}
where $t \in \mathbb{C}$ is the analytically-continued complex coupling, and $a$ is given as Eq.~\eqref{eq_coefficient_a}.
We show the resulting renormalization-group flows for $\mathcal{O}\,( t^5 )$ in Fig.~\ref{fig_Un}, together with the corresponding complex scaling dimensions.
Since the coefficient of the $\mathcal{O}\,( t^5 )$ term vanishes for $N=2$ irrespective of dimensions, the number of fixed points is reduced by one for $N=2$ compared with $N=1$ and $N=3$. 
Furthermore, for $N = 3$, we find a stable fixed point on the real axis with an anomalously large scaling dimension [Figs.~\ref{fig_Un}\,(d),\,(h),\,(l)], 
which seems to be an artifact of the perturbative renormalization group.

In one dimension [Figs.~\ref{fig_Un}\,(a)-(d)], all the fixed points are stable for each value of $N$.
For $N=1$, two complex-conjugate pairs of fixed points emerge, as in the case of the $\mathrm{O}\left(N\right)$ nonlinear sigma model. 
For $N=2$ and $N=3$, on the other hand, we only find a single complex-conjugate pair of fixed points, along with stable fixed points on the real axis.
For $N=2$, the stable fixed point on the negative real axis attracts flows from $t=-\infty$ and $t=0$. 
For $N=3$, the stable fixed points on the negative (positive) real axis attract flows from $t=-\infty$ ($t=\infty$) and $t=0$.
In two dimensions [Figs.~\ref{fig_Un}\,(e)-(h)], the complex fixed points have scaling dimensions with the positive real parts and thus form unstable spirals for all $N$.
This contrasts with the $\mathrm{O}\left(N\right)$ nonlinear sigma model, where the $N=1$ case yields stable complex fixed points.
In three dimensions [Figs.~\ref{fig_Un}\,(i)-(l)], all the complex fixed points form unstable spirals, as in the $\mathrm{O}\left(N\right)$ case. 
Moreover, on the real axis, two unstable fixed points appear for $N=1$, again similar to the $\mathrm{O}\left(N\right)$ case. 
For $N=3$, an unstable fixed point appears near the origin, and the renormalization-group flow approaches weak coupling $t=0$ for $0 \leq t < 0.177\cdots$, whereas it flows to the real stable fixed point $t=6.040\cdots$ for $t > 0.177\cdots$.

\begin{figure}[t]
\includegraphics[width=0.90\linewidth]{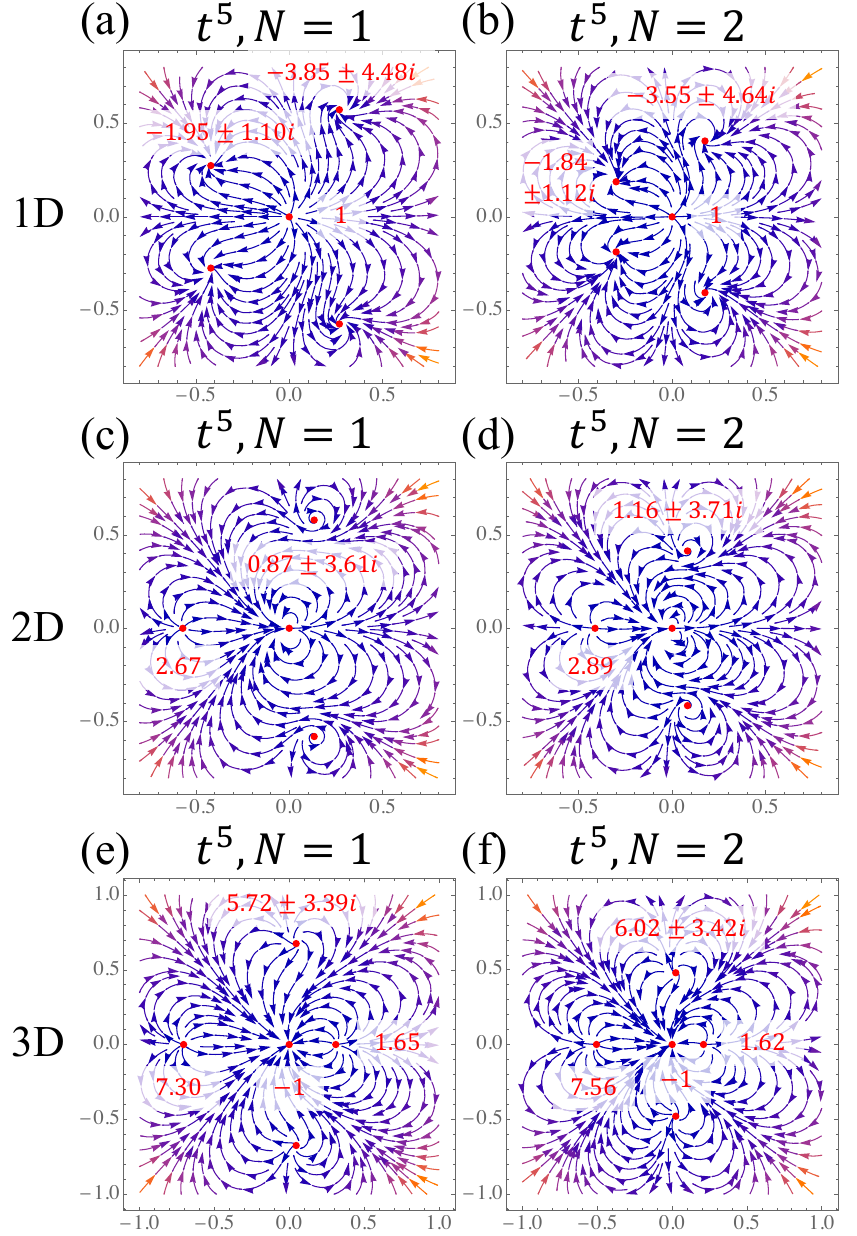}
\caption{Perturbative renormalization-group flow of the $\mathrm{Sp} \left( N \right)$ nonlinear sigma model in the complex-coupling plane at the quintic order in perturbation theory. 
Fixed points (red points) occur either on the real axis or in complex-conjugate pairs, and their scaling dimensions are also shown.
The top, middle, and bottom panels correspond to one, two, and three dimensions, respectively.
The first and second columns show the results for $N=1$ and $N=2$, respectively.}
    \label{fig_Spn}
\end{figure}

\subsection{\texorpdfstring{$\mathrm{Sp}\left(N\right)$}{Spn}}
\label{sec_Spn}

For the nonlinear sigma model whose target space is the symplectic group $\mathrm{Sp}\left(N\right)$, the renormalization-group equation up to $\mathcal O \left(t^5\right)$ in perturbation theory is given by~\cite{Hikami-81, Wegner-89}
\begin{align}
&\frac{dt}{dl} = \left( 2-d \right) t + \left( N+1 \right) t^2 + \frac{1}{2} \left( N+1 \right)^2 t^3 \notag\\
&\qquad\quad +\frac{3}{8} \left( N+1 \right)^3 t^4 + \frac{1}{8} \left( N+1 \right) c_1 \left( -2N \right) t^5,
\label{eq_Spn}
\end{align}
where $c_1\left(N\right)$ is defined in Eq.~\eqref{eq_coefficient_c1}. 
We complexify the coupling constant $t$ and investigate the renormalization-group flow in the complex plane at $\mathcal O\left(t^5\right)$, as shown in Fig.~\ref{fig_Spn}. 
As in the case of the $\mathrm O\left(N\right)$ nonlinear sigma model, up to $\mathcal O\left(t^4\right)$, the renormalization-group equation can be rewritten in a form independent of $N$ by rescaling $t$ as $x \coloneqq \left( N+1 \right) t$; 
when written as an equation for $x$, the expansion up to $\mathcal O\left(t^4\right)$ becomes identical to that of the $\mathrm{O} \left( N \right)$ case. 
At $\mathcal{O}\,( t^5 )$ in each dimension, we find qualitatively similar behavior for $N=1$ and $N=2$ (and likewise for $N>2$, not shown here).
This behavior differs from the $\mathrm O\left(N\right)$ nonlinear sigma model, where the $\mathrm O\left( 2\right)$ case yields purely radial or ill-defined renormalization-group flows.

In one dimension [Figs.~\ref{fig_Spn}\,(a),\,(b)], we find two complex-conjugate pairs of stable fixed points irrespective of the values of $N$, which is similar to the $\mathrm{O}\left(N\right)$ case for $N=1$.
In two dimensions [Figs.~\ref{fig_Spn}\:(c),\,(d)], a complex-conjugate pair of fixed points with unstable spirals emerges, and an unstable fixed point appears on the negative real axis for arbitrary $N$. 
This behavior contrasts with the $\mathrm{O} \left(N\right)$ case, which accompanies either stable or unstable complex fixed points in two dimensions depending on $N$. 
In addition, for $N=1$, no fixed point is present on the positive real axis, which also differs from the $\mathrm O\left(N\right)$ case. 
Finally, in three dimensions [Figs.~\ref{fig_Spn}\,(e),\,(f)], a complex-conjugate pair of fixed points with unstable spirals, together with two unstable fixed points on the real axis, appears regardless of $N$, closely resembling the $\mathrm{O}\left( 1 \right)$ case.

\begin{figure}[t]
\includegraphics[width=0.85\linewidth]{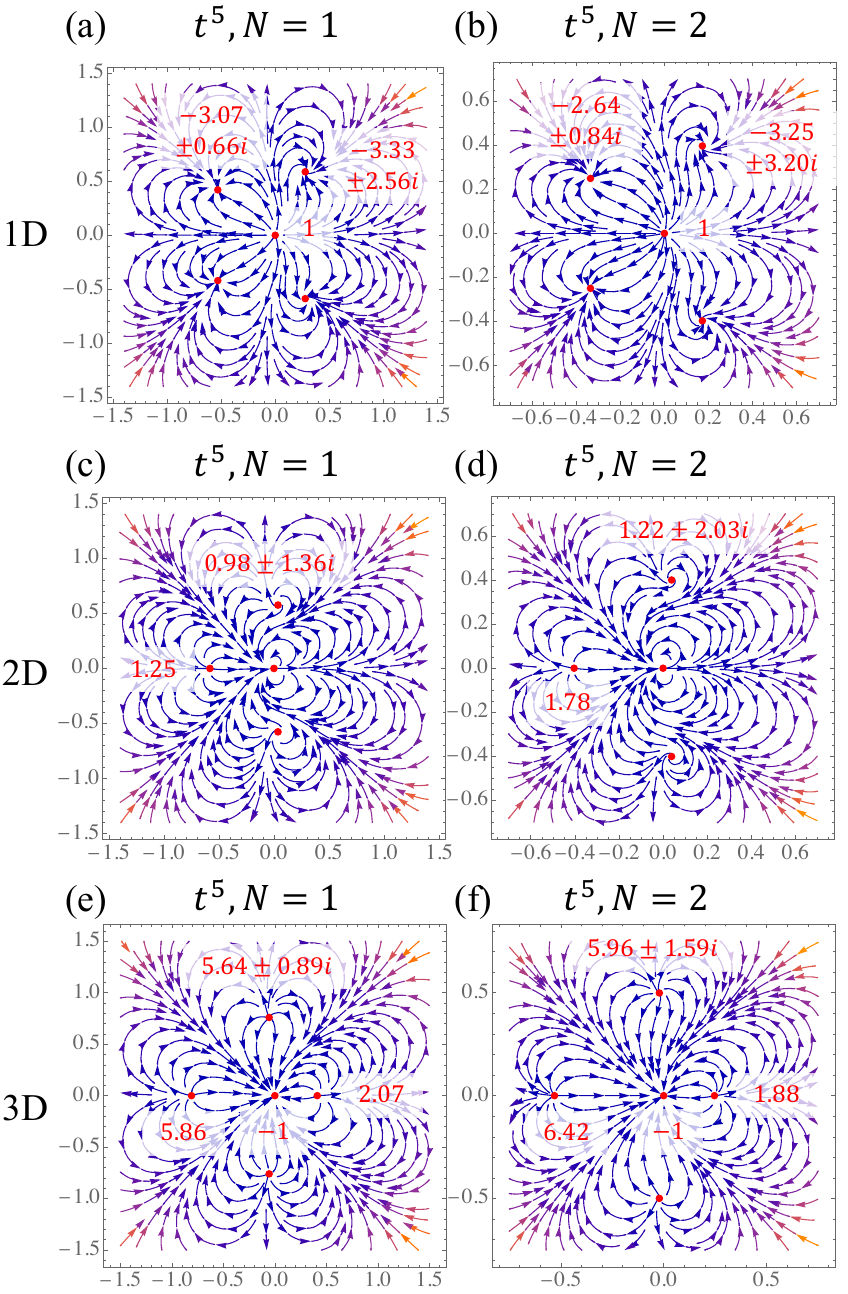}
\caption{Perturbative renormalization-group flow of the $\mathrm{U}\left( N \right)/\mathrm{O}\left( N \right)$ nonlinear sigma model in the complex-coupling plane at the quintic order in perturbation theory. 
Fixed points (red points) occur either on the real axis or in complex-conjugate pairs, and their scaling dimensions are also shown.
The top, middle, and bottom panels correspond to one, two, and three dimensions, respectively.
The first and second columns show the results for $N=1$ and $N=2$, respectively.}
    \label{fig_Un_div_On}
\end{figure}

\subsection{\texorpdfstring{$\mathrm{U}\left(N\right)/\mathrm{O}\left(N\right)$}{UndivOn}}

We investigate the renormalization-group flow in the complex-coupling plane for the $\mathrm{U}\left(N\right)/\mathrm{O}\left(N\right)$ nonlinear sigma model. 
The renormalization-group equation up to the fifth order in perturbation theory is given by~\cite{Hikami-81, Wegner-89}
\begin{align}
&\frac{dt}{dl} = \left( 2-d \right) t + N t^2 + \frac{N \left( N+2 \right)}{2} t^3\notag\\
&\qquad + \frac{N \left( 3N^2 + 10N + 8 \right)}{8} t^4 + \frac{N}{2}c_2\left( - \frac{N}{2} \right)t^5,
\label{eq_Un_div_On}
\end{align}
where
\begin{equation}
c_2 \left( N \right) \coloneqq -\frac{19}{3}N^3 + \left(\frac{43}{3}-8a\right)N^2 - \left( 9+8a \right) N + 1,
    \label{eq_coefficient_c2}
\end{equation}
with $a$ defined in Eq.~\eqref{eq_coefficient_a}.
We present the resulting renormalization-group flow, together with the associated complex scaling dimensions, in Fig.~\ref{fig_Un_div_On}.
In one, two, and three dimensions, we find no qualitative difference compared with the $\mathrm{Sp}\left(N\right)$ nonlinear sigma model discussed in Sec.~\ref{sec_Spn}. 

\begin{figure}[t]
\includegraphics[width=0.85\linewidth]{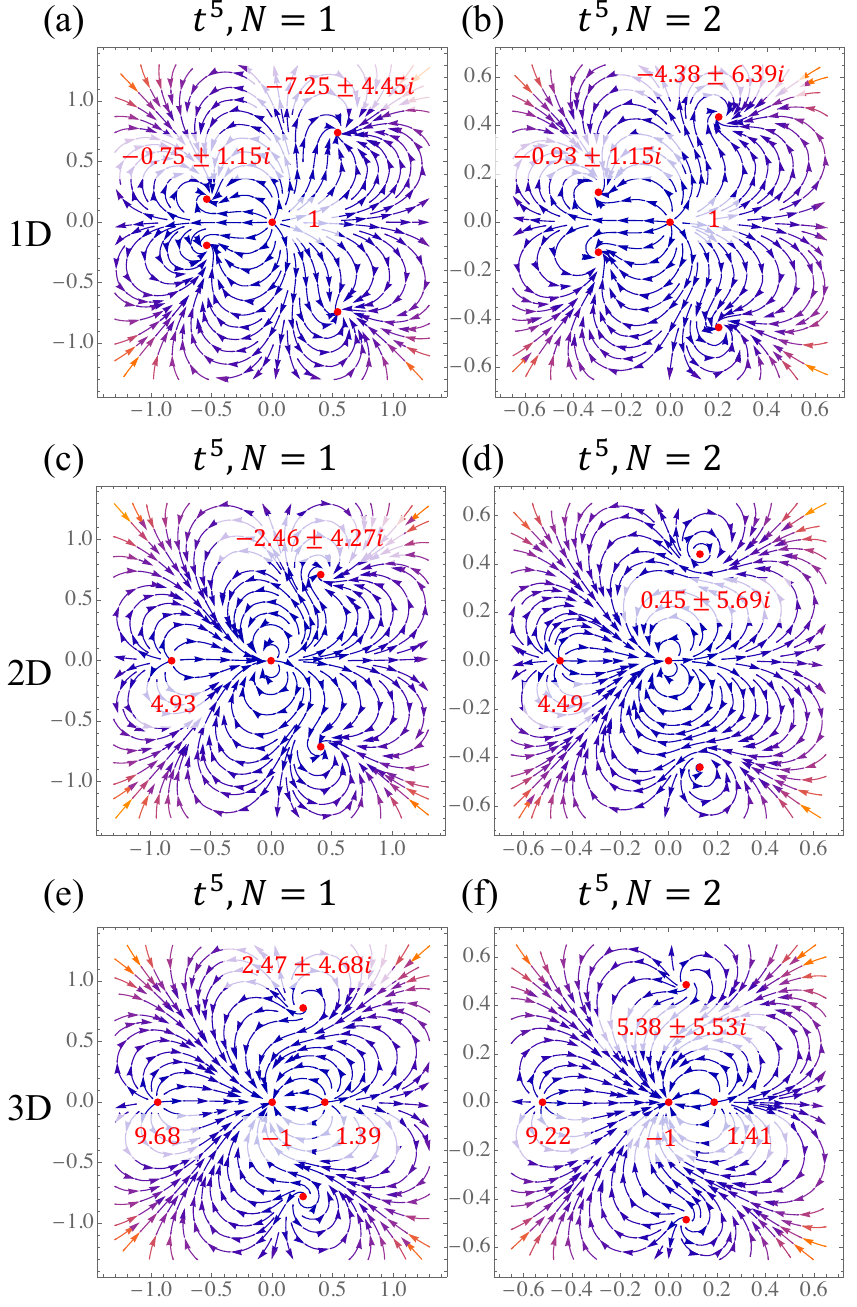}
\caption{Perturbative renormalization-group flow of the $\mathrm{U}\left( 2N \right)/\mathrm{Sp}\left( N \right)$ nonlinear sigma model in the complex-coupling plane at the quintic order in perturbation theory. 
Fixed points (red points) occur either on the real axis or in complex-conjugate pairs, and their scaling dimensions are also shown.
The top, middle, and bottom panels correspond to one, two, and three dimensions, respectively.
The first and second columns show the results for $N=1$ and $N=2$, respectively.
In two dimensions, the complex fixed points with the spiral renormalization-group flow are stable for $N=1$ but unstable for $N=2$.}
    \label{fig_U2n_div_Spn}
\end{figure}

\subsection{\texorpdfstring{$\mathrm{U}\left( 2N \right)/\mathrm{Sp}\left( N \right)$}{U2ndivSpn}}

The renormalization-group equation for the $\mathrm{U}\left( 2N \right)/\mathrm{Sp}\left( N \right)$ nonlinear sigma model is given by~\cite{Hikami-81, Wegner-89}
\begin{align}
&\frac{dt}{dl} = \left( 2-d \right) t + 2N t^2 + 2N \left( N-1\right) t^3 \notag\\
&\qquad\qquad + N \left( 3N^2-5N+2 \right) t^4 -N c_2 \left( N \right) t^5,
    \label{eq_U2n_div_Spn}
\end{align}
where $c_2\left(N\right)$ is defined as Eq.~\eqref{eq_coefficient_c2}.
We show the corresponding renormalization-group flows at $\mathcal{O}\,( t^5 )$ in Fig.~\ref{fig_U2n_div_Spn}.

In one [Figs.~\ref{fig_U2n_div_Spn}\,(a),\,(b)] and three [Figs.~\ref{fig_U2n_div_Spn}\,(e),\,(f)] dimensions, the renormalization-group flow exhibits behavior similar to the $\mathrm{Sp}\left(N\right)$ case in Sec.~\ref{sec_Spn}.
In two dimensions [Figs.~\ref{fig_U2n_div_Spn}\,(c),\,(d)], for any $N$, a complex-conjugate pair of fixed points occurs, together with an unstable fixed point on the negative real axis, which arises only for $N>2$ in the $\mathrm{O}\left(N\right)$ case. 
For $N=1$, the complex fixed points have the negative real part of scaling dimensions and are hence stable. 
For $N=2$ (and likewise for $N>2$, not shown), on the other hand, this pair of complex fixed points forms unstable spirals, indicating criticality unique to complexified field theory. 

\begin{figure}[t]
\includegraphics[width=0.87\linewidth]{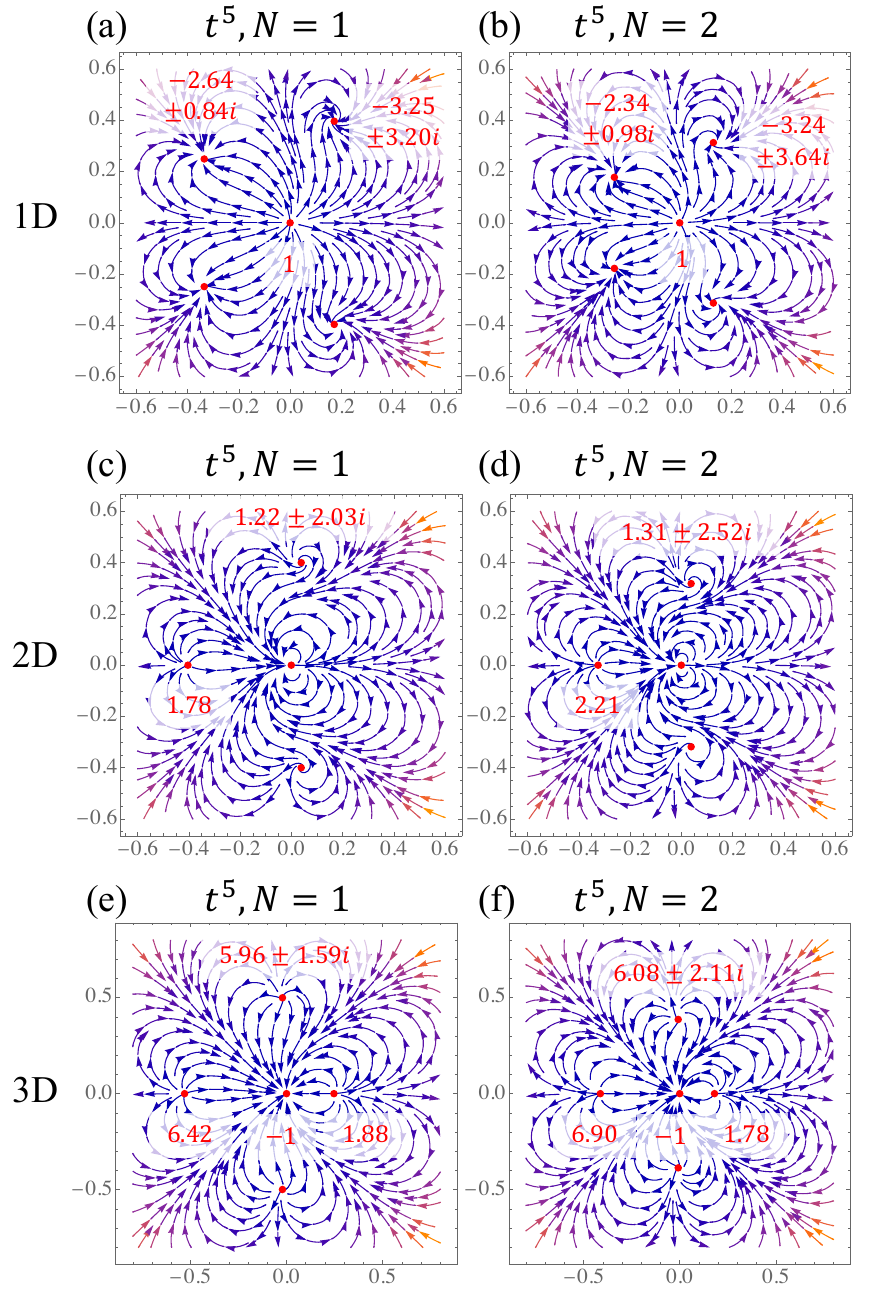}
\caption{Perturbative renormalization-group flow of the $\mathrm{Sp}\left( N \right)/\mathrm{U}\left( N \right)$ nonlinear sigma model in the complex-coupling plane at the quintic order in perturbation theory. 
Fixed points (red points) occur either on the real axis or in complex-conjugate pairs, and their scaling dimensions are also shown.
The top, middle, and bottom panels correspond to one, two, and three dimensions, respectively.
The first and second columns show the results for $N=1$ and $N=2$, respectively.}
    \label{fig_Spn_div_Un}
\end{figure}

\subsection{\texorpdfstring{$\mathrm{Sp}\left( N \right)/\mathrm{U}\left( N \right)$}{SpndivUn}}

The renormalization-group equation for the $\mathrm{Sp}\left( N \right)/\mathrm{U}\left( N \right)$ nonlinear sigma model is given by~\cite{Hikami-81, Wegner-89}
\begin{align}
&\frac{dt}{dl} = \left( 2-d \right) t + \left( N+1 \right) t^2 + \frac{N^2+3N+4}{2} t^3\notag\\
&\qquad +\frac{3N^3+14N^2+35N+28}{8} t^4-c_3 \left( N \right) t^5,
    \label{eq_Spn_div_Un}
\end{align}
with
\begin{equation}
c_3\left(N\right)\coloneqq - \frac{19}{48}N^4 - \frac{119}{48}N^3 - \frac{380}{48}N^2 - \frac{578}{48}N - \frac{376}{48} .
    \label{eq_coefficient_c3}
\end{equation}
We present the corresponding renormalization-group flows in the complex-coupling plane in Fig.~\ref{fig_Spn_div_Un}.
We find qualitatively similar behavior to the $\mathrm{Sp}\left(N\right)$ nonlinear sigma model discussed in Sec.~\ref{sec_Spn}. 

\begin{figure*}[t]
\includegraphics[width=0.8\linewidth]{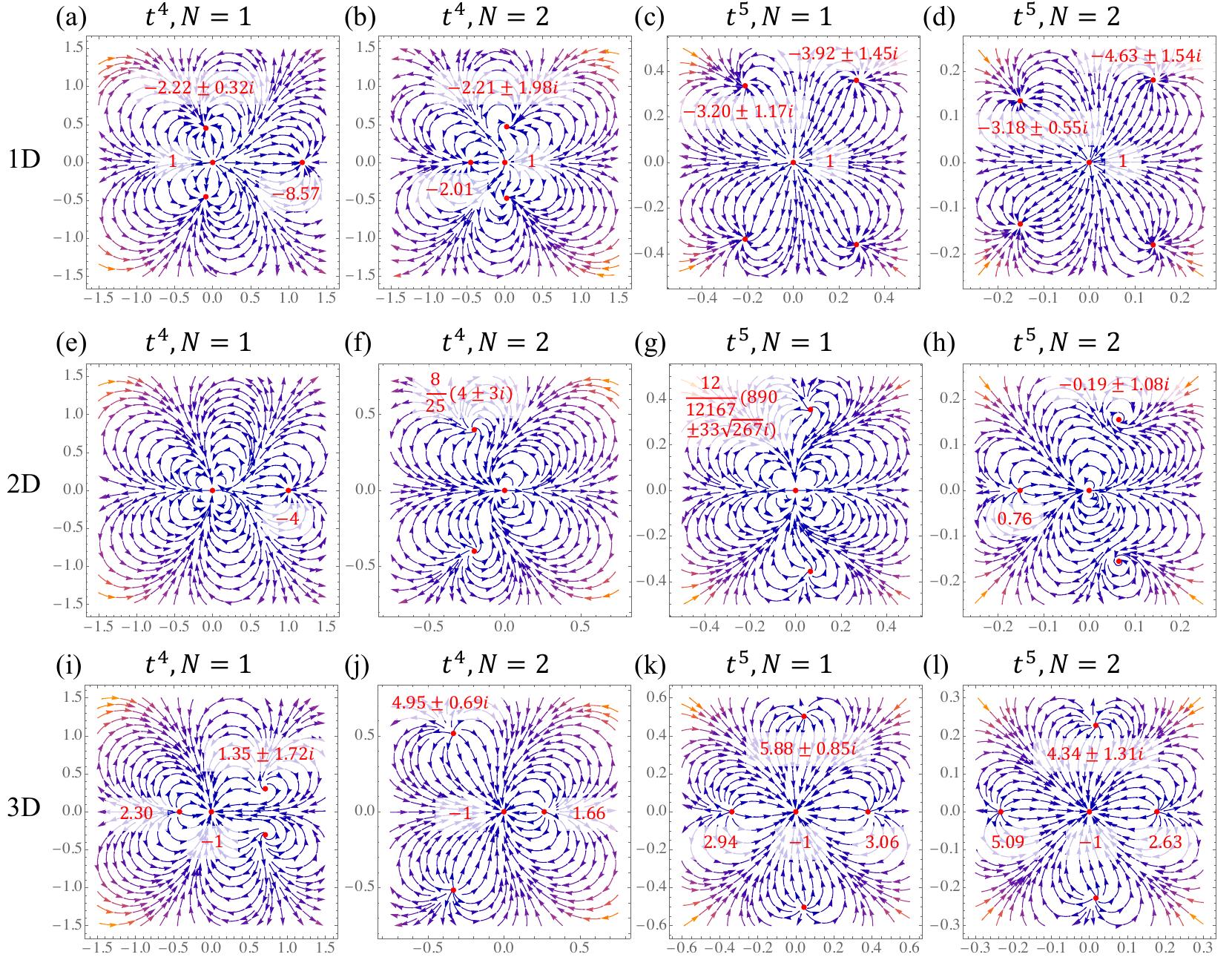}
\caption{Perturbative renormalization-group flow of the $\mathrm{O}\left( 2N \right)/\mathrm{U}\left( N \right)$ nonlinear sigma model in the complex-coupling plane. 
Fixed points (red points) occur either on the real axis or in complex-conjugate pairs, and their scaling dimensions are also shown.
The top, middle, and bottom panels correspond to one, two, and three dimensions, respectively. 
The first and second (third and fourth) columns show the results at the quartic (quintic) order in perturbation theory for $N=1$ and $N=2$, respectively. 
In two dimensions for $N=2$, the complex fixed points are unstable up to the quartic order but become stable at the quintic order.}
\label{fig_O2n_div_Un}
\end{figure*}

\subsection{\texorpdfstring{$\mathrm{O}\left( 2N \right)/\mathrm{U}\left( N \right)$}{O2ndivUn}}

The renormalization-group equation for the $\mathrm{O}\left( 2N \right)/\mathrm{U}\left( N \right)$ nonlinear sigma model is given by~\cite{Hikami-81, Wegner-89, TongWang-23}
\begin{align}
&\frac{dt}{dl} = \left( 2-d \right) t + \left( 2N-2 \right) t^2 + \left( 2N^2-6N+8 \right) t^3 \notag\\
&+ \left( 3N^3-14N^2+35N-28 \right) t^4 - 16 c_3 \left(-2N\right) t^5,
    \label{eq_O2n_div_Un}
\end{align}
with $c_3$ defined in Eq.~\eqref{eq_coefficient_c3}.
We investigate the corresponding renormalization-group flow at $\mathcal{O}\,( t^5 )$ in the complex-coupling plane, as shown in Fig.~\ref{fig_O2n_div_Un}.

At $\mathcal{O}\,( t^5 )$ in one [Figs.~\ref{fig_O2n_div_Un}\,(c),\,(d)] and three [Figs.~\ref{fig_O2n_div_Un}\,(k),\,(l)] dimensions, we find qualitatively similar renormalization-group flows to the $\mathrm{Sp}\left(N\right)$ nonlinear sigma model discussed in Sec.~\ref{sec_Spn}.
In two dimensions, 
for $N=1$, both the linear and quadratic terms in Eq.~\eqref{eq_O2n_div_Un} vanish, and the number of fixed points is reduced to $n-2$ when the perturbation theory is truncated at $\mathcal{O}\,( t^n )$.
Then, the complex fixed points forming unstable spirals emerge only at $\mathcal{O}\,( t^5 )$.
For $N=2$, on the other hand, a complex-conjugate pair of fixed points emerges already at $\mathcal{O}\,( t^4 )$ and persists at $\mathcal{O}\,( t^5 )$.
However, whereas the complex fixed points at $\mathcal{O}\,( t^4 )$ are unstable, the corresponding fixed points at $\mathcal{O}\,( t^5 )$ become stable.

\begin{figure}[t]
\includegraphics[width=0.85\linewidth]{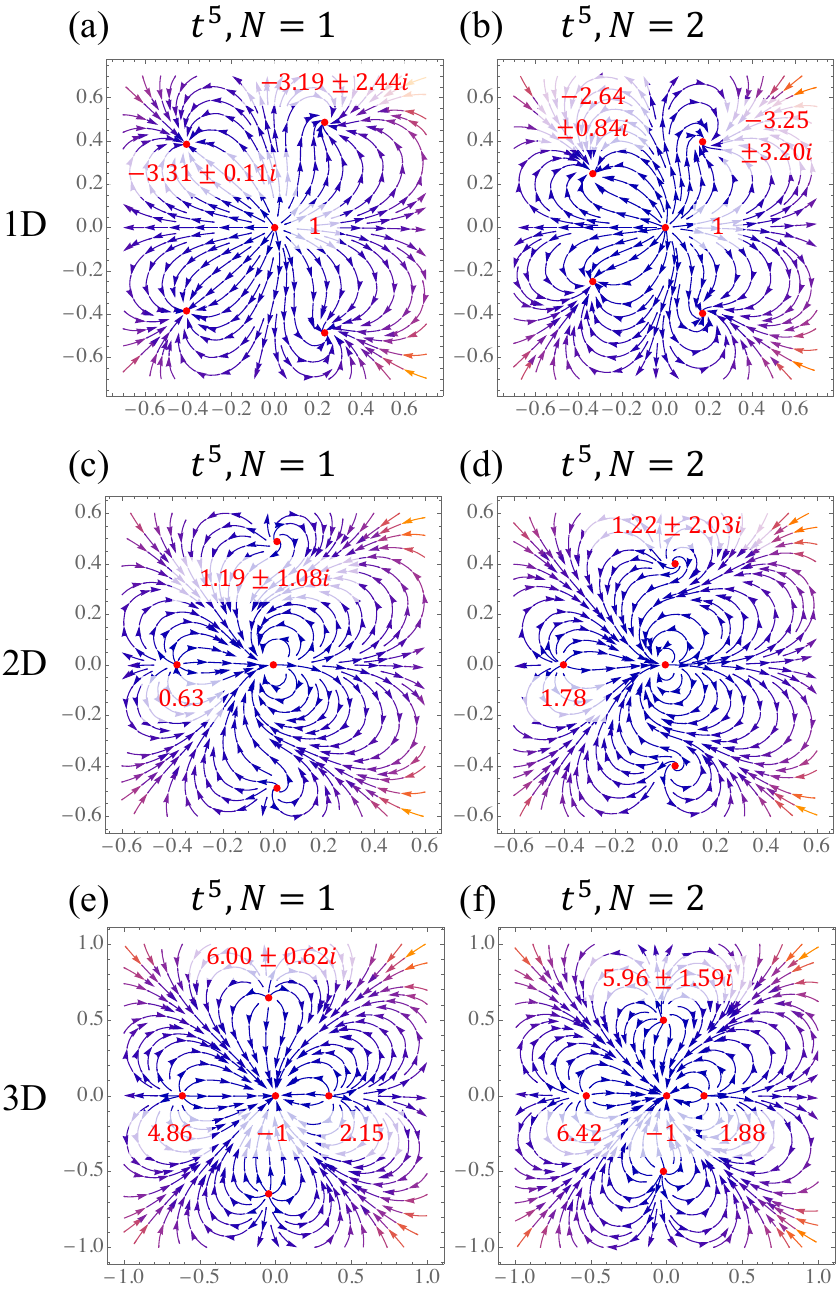}
\caption{Perturbative renormalization-group flow of the $\mathrm{U}\left( N \right)/\mathrm{U}\left( p \right)\times\mathrm{U}\left( N-p \right)$ nonlinear sigma model with $p=N/2$ in the complex-coupling plane at the quintic order in perturbation theory. 
Fixed points (red points) occur either on the real axis or in complex-conjugate pairs, and their scaling dimensions are also shown.
The top, middle, and bottom panels correspond to one, two, and three dimensions, respectively.
The first and second columns show the results for $N=1$ and $N=2$, respectively.}
    \label{fig_Un_div_UptimesUnmp}
\end{figure}

\subsection{\texorpdfstring{$\mathrm{U}\left( N \right)/\mathrm{U}\left( p \right)\times\mathrm{U}\left( N-p \right)$}{UndivUptimesUnmp}}

The renormalization-group equation for the $\mathrm{U}\left( N \right)/\mathrm{U}\left( p \right)\times\mathrm{U}\left( N-p \right)$ nonlinear sigma model is given by~\cite{Hikami-81, Wegner-89}
\begin{align}
&\frac{dt}{dl} = \left( 2-d \right) t + N t^2 + 2 \left[ 1 + p \left( N-p \right) \right] t^3\notag\\
&\qquad\qquad +\frac{N \left[ 3p \left( N-p \right) + 7 \right]}{2}t^4-c_4 \left( N,p \right) t^5,
    \label{eq_Un_div_UptimesUnmp}
\end{align}
with
\begin{align}
&c_4 \left(N,p\right) \coloneqq-\frac{1}{3} p N^2 \left( N - p \right) - 5 p^2 \left( N - p \right)^2 \notag\\
&\qquad\qquad\qquad\qquad- \frac{11}{6}N^2 - 11p \left( N - p \right) - 6.
\end{align}
For $p=N/2$, we investigate the corresponding renormalization-group flow in the complex-coupling plane, as shown in Fig.~\ref{fig_Un_div_UptimesUnmp}.
It is qualitatively similar to that of the $\mathrm{Sp}\left(N\right)$ nonlinear sigma model discussed in Sec.~\ref{sec_Spn}. 

\begin{figure}[t]
\includegraphics[width=1.0\linewidth]{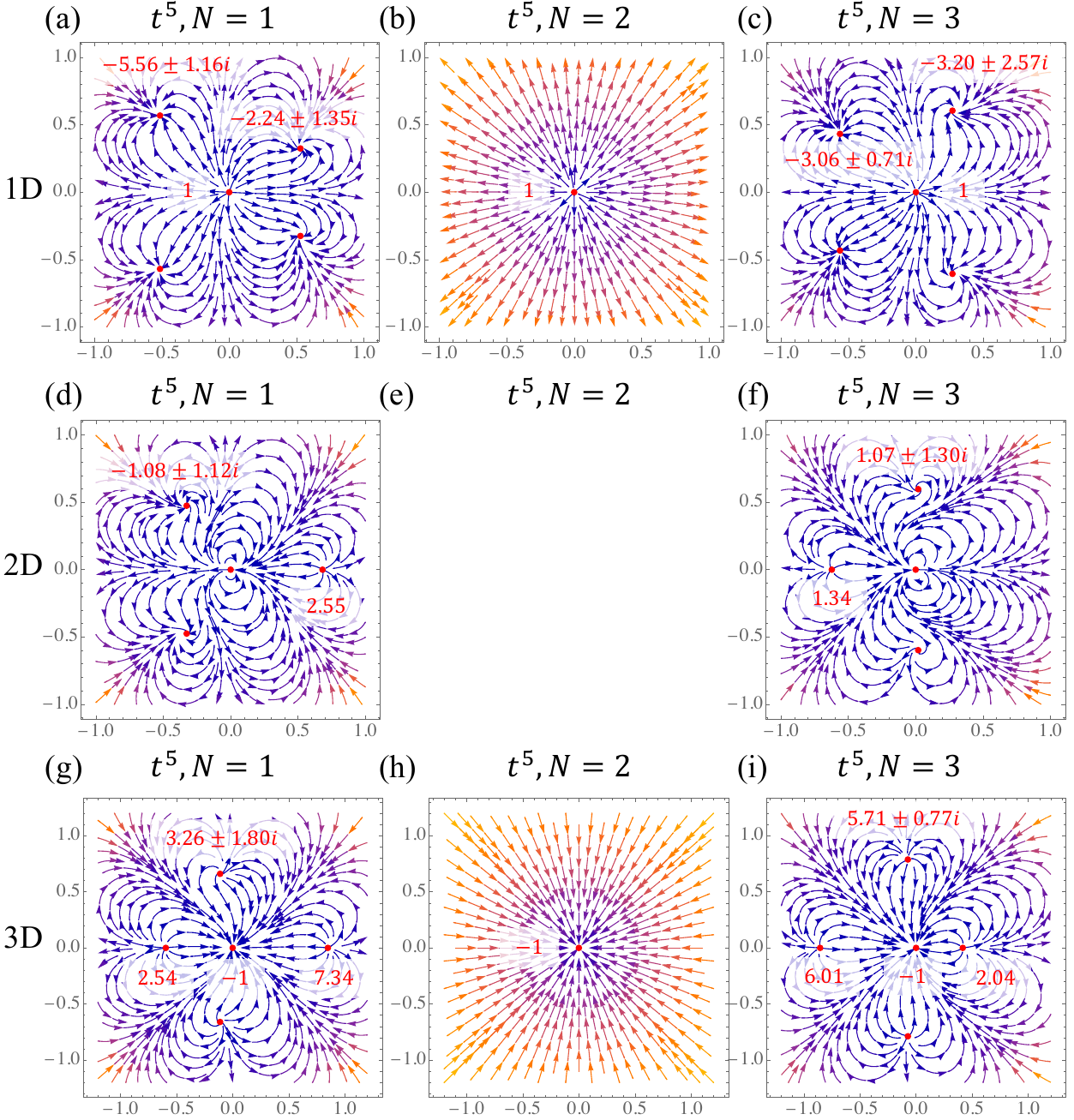}
\caption{Perturbative renormalization-group flow of the $\mathrm{O}\left( N \right)/\mathrm{O}\left( p \right)\times\mathrm{O}\left( N- p \right)$ nonlinear sigma model with $p=N/2$ in the complex-coupling plane at the quintic order in perturbation theory. 
Fixed points (red points) occur either on the real axis or in complex-conjugate pairs, and their scaling dimensions are also shown.
The top, middle, and bottom panels correspond to one, two, and three dimensions, respectively.
The first, second, and third columns show the results for $N=1$, $N=2$, and $N=3$, respectively.}
    \label{fig_On_div_OptimesOnmp}
\end{figure}

\subsection{\texorpdfstring{$\mathrm{O}\left( N \right)/\mathrm{O}\left( p \right)\times\mathrm{O}\left( N-p \right)$}{OndivOptimesOnmp}}

The renormalization-group equation for the $\mathrm{O}\left( N \right)/\mathrm{O}\left( p \right)\times\mathrm{O}\left( N-p \right)$ nonlinear sigma model is given by~\cite{Hikami-81, Wegner-89}
\begin{align}
&\frac{dt}{dl} = \left( 2-d \right) t + \left( N-2 \right) t^2 + \left[ 2p \left( N-p \right) - N\right] t^3\notag\\
&+\left[\frac{3}{2} p N \left( N-p \right)-\frac{5}{4} N^2 + p \left(N-p \right) + \frac{N}{2}\right] t^4\notag\\
&-c_5\left(N,p\right)t^5,
    \label{eq_On_div_OptimesOnmp}
\end{align}
where
\begin{align}
&c_5 \left(N,p\right) \coloneqq  - \frac{1}{3}pN^2 \left( N - p \right) - 5p^2 \left( N - p \right)^2 + \frac{5}{12}N^3 \notag\\
&+ \left(\frac{23}{6} + 8a\right) p N \left( N - p \right) - \left(-\frac{2}{3} + 16a\right) p \left( N - p \right) \notag\\
&- \left(\frac{7}{6} + 16a\right)N^2 - \left(\frac{1}{3} - 64a\right)N + 64a,
\end{align}
with $a$ defined as Eq.~\eqref{eq_coefficient_a}.
For $p=N/2$, we study the resulting renormalization-group flow at $\mathcal{O}\,( t^5 )$, including the complex scaling dimensions, as shown in Fig.~\ref{fig_On_div_OptimesOnmp}.
In one, two, and three dimensions, we find no qualitative difference in the renormalization-group behavior compared with the $\mathrm{O}\left(N\right)$ nonlinear sigma model discussed in Sec.~\ref{sec: On}. 

\begin{figure}[t]
\includegraphics[width=1.0\linewidth]{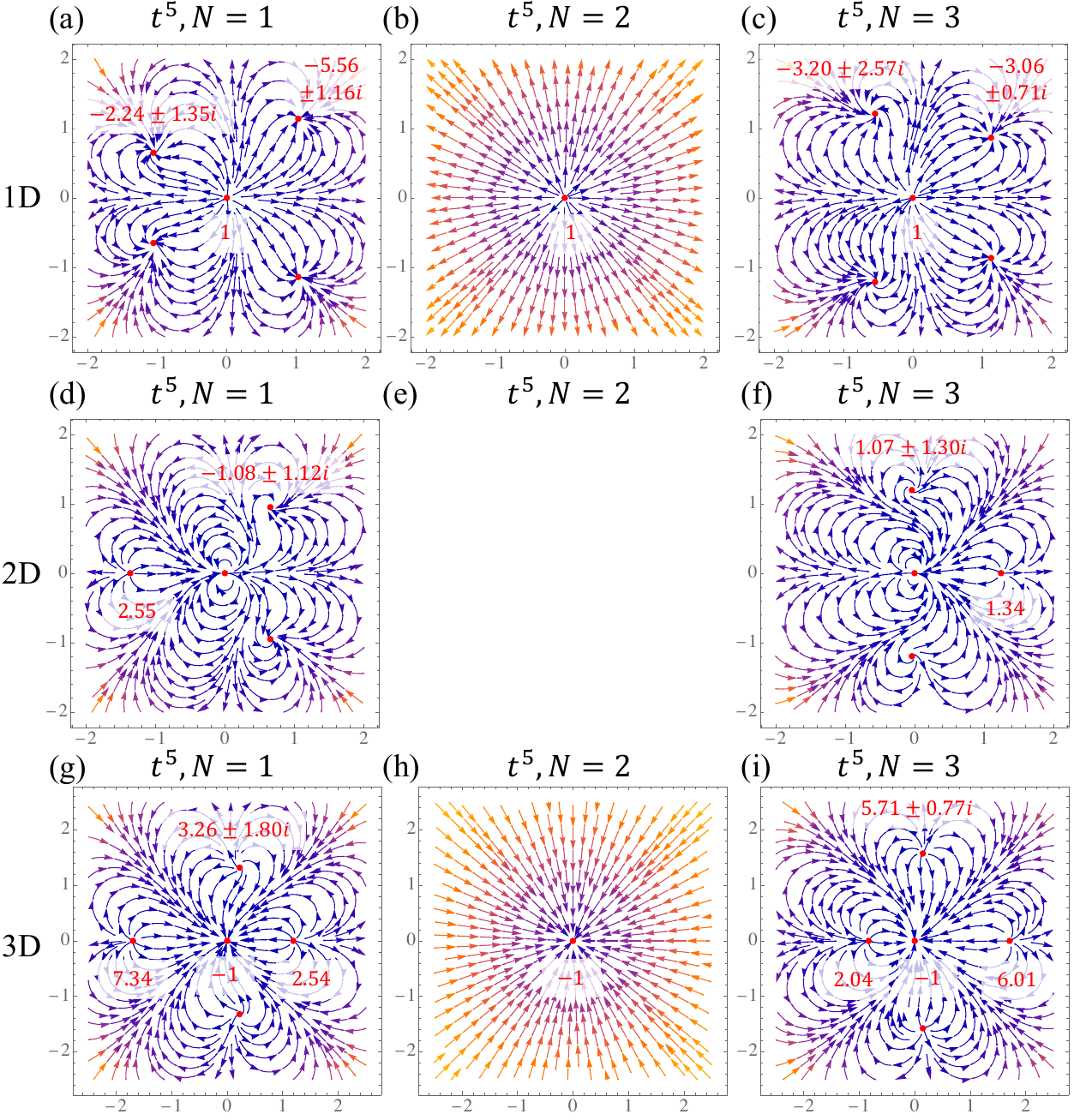}
\caption{Perturbative renormalization-group flow of the $\mathrm{Sp}\left( N \right)/\mathrm{Sp}\left( p \right)\times\mathrm{Sp}\left( N- p \right)$ nonlinear sigma model with $p=N/2$ in the complex-coupling plane at the quintic order in perturbation theory. 
Fixed points (red points) occur either on the real axis or in complex-conjugate pairs, and their scaling dimensions are also shown.
The top, middle, and bottom panels correspond to one, two, and three dimensions, respectively.
The first, second, and third columns show the results for $N=1$, $N=2$, and $N=3$, respectively.
Compared with Fig.~\ref{fig_On_div_OptimesOnmp}, the renormalization-group flow is reversed with respect to the origin after an appropriate rescaling.}
    \label{fig_Spn_div_SpptimesSpnmp}
\end{figure}

\subsection{\texorpdfstring{$\mathrm{Sp}\left( N \right)/\mathrm{Sp}\left( p \right)\times\mathrm{Sp}\left( N- p \right)$}{SpndivSpptimesSpnmp}}

The renormalization-group equation for the $\mathrm{Sp}\left( N \right)/\mathrm{Sp}\left( p \right)\times\mathrm{Sp}\left( N- p \right)$ nonlinear sigma model is given by~\cite{Hikami-81, Wegner-89}
\begin{align}
&\frac{dt}{dl}= \left( 2-d \right) t -\frac{1}{2} \left( N-2 \right) t^2 + \frac{1}{4}\left[ 2p \left( N-p \right) - N \right]t^3\notag\\
&-\frac{1}{8}\left[\frac{3}{2}pN \left( N-p \right)-\frac{5}{4}N^2+p \left( N-p \right)+\frac{N}{2} \right]t^4\notag\\
&-\frac{c_5\left(N,p\right)}{16}t^5.
    \label{eq_Spn_div_SpptimesSpnmp}
\end{align}
Notably, Eq.~\eqref{eq_Spn_div_SpptimesSpnmp} is mapped to Eq.~\eqref{eq_On_div_OptimesOnmp} by the replacement $t \mapsto -t/2$, 
which reflects the duality between the corresponding symmetric spaces.
Consequently, the renormalization-group flow of Eq.~\eqref{eq_Spn_div_SpptimesSpnmp}, shown in Fig.~\ref{fig_Spn_div_SpptimesSpnmp}, coincides with that in Fig.~\ref{fig_On_div_OptimesOnmp} up to a reversal of direction.

\section{Discussion}
    \label{sec: conclusion}

We have shown that complex nonlinear sigma models generically exhibit fixed points characterized by complex scaling dimensions and critical exponents, 
inducing the spiral renormalization-group flows.
We have also compared the complex critical exponents obtained from our perturbative renormalization-group analysis with the corresponding exact values.
While extending such a comparative analysis to three dimensions is of considerable significance, 
exact results of complex scaling dimensions and critical exponents are unavailable in three dimensions.
In this respect, it is noteworthy that numerical investigations of scaling dimensions have recently been conducted for the Yang-Lee nonunitary conformal field theory in three dimensions~\cite{Fan-25, Cruz-25, Miro-25}.
Furthermore, regarding the two-dimensional $\mathrm{O} \left( 2 \right)$ nonlinear sigma model, the perturbative renormalization-group equation~\eqref{eq_On} vanishes, and the critical behavior is instead governed by the Berezinskii-Kosterlitz-Thouless transition~\cite{Berezinskii-71, *Berezinskii-72, Kosterlitz-Thouless-73, *Kosterlitz-74}.
It remains unclear whether the universal properties of the Berezinskii-Kosterlitz-Thouless transition are changed upon complexifying the coupling constants.

Recently, the complex $\mathrm{O} \left( 3 \right)$ nonlinear sigma model has been shown to emerge in a one-dimensional quantum antiferromagnet with non-Hermitian perturbations~\cite{Yang-26}.
It merits further research to investigate microscopic realizations of other complex nonlinear sigma models.
In general, non-Hermitian Hamiltonians describe the no-jump dynamics of quantum master equations and more generally individual quantum trajectories, corresponding to the nonunitary quantum dynamics under measurement~\cite{Plenio-review, Daley-review}.
We expect that the complex nonlinear sigma models studied in this work can appear as effective descriptions of such monitored quantum dynamics.
It is also an important direction for future work to study whether they arise at the level of quantum master equations, for example, through the Schwinger-Keldysh formalism.
Although disordered free fermions, both at equilibrium~\cite{Wegner-79, *Schafer-80, Gorkov-79, Hikami-80, Efetov-80, Altshuler-80} and far from equilibrium~\cite{Jian-22, *Jian-23, Yang-23, Fava-23, Poboiko-23}, are effectively captured by nonlinear sigma models, the coupling constant is usually constrained to real values.
It therefore remains an open question whether complex nonlinear sigma models can arise in similar settings.

\begin{acknowledgments}
K.Y. is supported by JSPS Program for Forming Japan's Peak Research Universities (J-PEAKS) Grant No.~JPJS00420230008,
KAKENHI Grant No.~JP25K17327,
Hirose Foundation,
Precise Measurement Technology Promotion Foundation,
Fujikura Foundation,
Toyota Riken Scholar Program,
and Support Center for Advanced Telecommunications Technology Research.
K.K. is supported by JSPS KAKENHI Grant No.~JP24H00945, No.~JP26H02015, No.~JP26K06970, and No.~JP26K17046, and JST FOREST Program Grant No.~JPMJFR256P.
\end{acknowledgments}

\appendix

\section{Borel-Pad\'e resummation}
    \label{asec: Borel-Pade}

We summarize the Borel-Pad\'e resummation procedure for obtaining the critical exponent.
Its application to the conventional nonlinear sigma models is found, for example, in Refs.~\cite{Ueoka-Slevin-14, *Ueoka-Slevin-17, TongWang-23}.
We here focus on the beta function in Eq.~\eqref{eq_On} for $d=2$. 
We introduce the polynomial
\begin{equation}
    f \left( t \right) \coloneqq \frac{\beta \left( t \right)}{t^2},
\end{equation}
and apply the Borel-Pad\'e resummation to it.
Specifically, for
\begin{equation}    
    f \left( t \right) = \sum_{k=0}^l f_k t^k,
\end{equation}
the corresponding Borel transform is
\begin{equation}
    \tilde{f} \left( t \right) = \sum_{k=0}^l \frac{f_k}{k!} t^k.
\end{equation}
Here, the functions $f$ and $\tilde{f}$ are related by
\begin{align}
    f \left( t \right) 
    &= \int_0^{\infty} e^{-s} \tilde{f} \left( st \right) ds \nonumber \\
    &= \frac{1}{t} \int_{\mathcal{C}_\theta} e^{-u/t} \tilde{f} \left( u \right) du,
        \label{aeq: h-tildeh}
\end{align}
where the integration contour $\mathcal{C}_\theta$ is given as $\mathcal{C}_\theta \coloneqq \{ e^{i\theta} r\,|\,r \geq 0\}$ with $\theta \coloneqq \arg t$.

Then, we approximate $\tilde{f}$ by a rational function,
\begin{equation}
    \tilde{f} \left( t \right) \simeq \frac{p \left( t \right)}{q \left( t \right)},
\end{equation}
where $p$ and $q$ are polynomials of order $m$ and $n$, respectively, satisfying $m+n=l$.
When we additionally impose $m < n$, we have
\begin{equation}
    \lim_{t\to\infty} \tilde{f} \left( t \right) \simeq \lim_{t\to\infty} \frac{p \left( t \right)}{q \left( t \right)} = 0.
\end{equation}
In our calculations, we employ the Pad\'e approximants $[m, n]=[0, 3], [1, 2]$.

Furthermore, we decompose the rational function $p/q$ into partial fractions as
\begin{equation}
    \frac{p \left( t \right)}{q \left( t \right)} = \sum_{i=1}^{n} \frac{a_i}{t-\lambda_i} \quad \left( a_i, \lambda_i \in \mathbb{C} \right),
\end{equation}
where $\lambda_i$'s are the roots of the polynomial $q$.
Substituting this decomposition into Eq.~\eqref{aeq: h-tildeh}, we obtain the Borel-Pad\'e approximation $F$ of the series $f$ as,
\begin{equation}
    F \left( t \right) = \frac{1}{t} \sum_{i=1}^{n} a_i B \left( \frac{\lambda_i}{t} \right),
\end{equation}
where the function $B$ is defined as
\begin{equation}
    B \left( x \right) \coloneqq 
    e^{-x} E_1 \left( -x \right) \quad \left( x \in \mathbb{C} \backslash \left[ 0, \infty \right) \right),
\end{equation}
with the exponential integral 
\begin{align}
    E_1 \left( x \right) &\coloneqq \int_{x}^{\infty} \frac{e^{-t}}{t} dt.
\end{align}
Here, if a Pad\'e pole lies on $\mathcal{C}_\theta$, the contour is deformed so as to pass infinitesimally above or below the pole.
Moreover, $E_1 \left( x \right)$ is evaluated on its principal branch, with the branch cut along the negative real axis $\left( -\infty, 0 \right]$.
Finally, the beta function is approximated as 
\begin{equation}
    \beta \left( t \right) \simeq t^2 F \left( t \right).
\end{equation}

It is notable that the Borel-Pad\'e resummation employed in Refs.~\cite{Ueoka-Slevin-14, *Ueoka-Slevin-17, TongWang-23} is implemented so as to satisfy
\begin{equation}
    \lim_{d\to \infty} \nu = \frac{1}{2}.
\end{equation}
However, this constraint is not necessarily applicable to the complex fixed points investigated in the present work.
We thus do not impose it in our Borel-Pad\'e resummation procedure.




\begin{thebibliography}{94}%
\makeatletter
\providecommand \@ifxundefined [1]{%
 \@ifx{#1\undefined}
}%
\providecommand \@ifnum [1]{%
 \ifnum #1\expandafter \@firstoftwo
 \else \expandafter \@secondoftwo
 \fi
}%
\providecommand \@ifx [1]{%
 \ifx #1\expandafter \@firstoftwo
 \else \expandafter \@secondoftwo
 \fi
}%
\providecommand \natexlab [1]{#1}%
\providecommand \enquote  [1]{``#1''}%
\providecommand \bibnamefont  [1]{#1}%
\providecommand \bibfnamefont [1]{#1}%
\providecommand \citenamefont [1]{#1}%
\providecommand \href@noop [0]{\@secondoftwo}%
\providecommand \href [0]{\begingroup \@sanitize@url \@href}%
\providecommand \@href[1]{\@@startlink{#1}\@@href}%
\providecommand \@@href[1]{\endgroup#1\@@endlink}%
\providecommand \@sanitize@url [0]{\catcode `\\12\catcode `\$12\catcode
  `\&12\catcode `\#12\catcode `\^12\catcode `\_12\catcode `\%12\relax}%
\providecommand \@@startlink[1]{}%
\providecommand \@@endlink[0]{}%
\providecommand \url  [0]{\begingroup\@sanitize@url \@url }%
\providecommand \@url [1]{\endgroup\@href {#1}{\urlprefix }}%
\providecommand \urlprefix  [0]{URL }%
\providecommand \Eprint [0]{\href }%
\providecommand \doibase [0]{https://doi.org/}%
\providecommand \selectlanguage [0]{\@gobble}%
\providecommand \bibinfo  [0]{\@secondoftwo}%
\providecommand \bibfield  [0]{\@secondoftwo}%
\providecommand \translation [1]{[#1]}%
\providecommand \BibitemOpen [0]{}%
\providecommand \bibitemStop [0]{}%
\providecommand \bibitemNoStop [0]{.\EOS\space}%
\providecommand \EOS [0]{\spacefactor3000\relax}%
\providecommand \BibitemShut  [1]{\csname bibitem#1\endcsname}%
\let\auto@bib@innerbib\@empty
\bibitem [{\citenamefont {Goldenfeld}(1992)}]{Goldenfeld-textbook}%
  \BibitemOpen
  \bibfield  {author} {\bibinfo {author} {\bibfnamefont {N.}~\bibnamefont
  {Goldenfeld}},\ }\href {https://doi.org/10.1201/9780429493492} {\emph
  {\bibinfo {title} {{Lectures on Phase Transitions and the Renormalization
  Group}}}}\ (\bibinfo  {publisher} {Westview Press},\ \bibinfo {address}
  {Boulder},\ \bibinfo {year} {1992})\BibitemShut {NoStop}%
\bibitem [{\citenamefont {Cardy}(1996)}]{Cardy-textbook}%
  \BibitemOpen
  \bibfield  {author} {\bibinfo {author} {\bibfnamefont {J.}~\bibnamefont
  {Cardy}},\ }\href {https://doi.org/10.1017/CBO9781316036440} {\emph {\bibinfo
  {title} {{Scaling and Renormalization in Statistical Physics}}}}\ (\bibinfo
  {publisher} {Cambridge University Press},\ \bibinfo {address} {Cambridge,
  England},\ \bibinfo {year} {1996})\BibitemShut {NoStop}%
\bibitem [{\citenamefont {Sachdev}(1999)}]{Sachdev-textbook}%
  \BibitemOpen
  \bibfield  {author} {\bibinfo {author} {\bibfnamefont {S.}~\bibnamefont
  {Sachdev}},\ }\href {https://doi.org/10.1017/CBO9780511973765} {\emph
  {\bibinfo {title} {{Quantum Phase Transitions}}}}\ (\bibinfo  {publisher}
  {Cambridge University Press},\ \bibinfo {address} {Cambridge, England},\
  \bibinfo {year} {1999})\BibitemShut {NoStop}%
\bibitem [{\citenamefont {Zirnbauer}(1996)}]{Zirnbauer-96}%
  \BibitemOpen
  \bibfield  {author} {\bibinfo {author} {\bibfnamefont {M.~R.}\ \bibnamefont
  {Zirnbauer}},\ }\bibfield  {title} {\bibinfo {title} {{Riemannian symmetric
  superspaces and their origin in random‐matrix theory}},\ }\href
  {https://doi.org/10.1063/1.531675} {\bibfield  {journal} {\bibinfo  {journal}
  {J. Math. Phys.}\ }\textbf {\bibinfo {volume} {37}},\ \bibinfo {pages} {4986}
  (\bibinfo {year} {1996})}\BibitemShut {NoStop}%
\bibitem [{\citenamefont {Altland}\ and\ \citenamefont
  {Zirnbauer}(1997)}]{AZ-97}%
  \BibitemOpen
  \bibfield  {author} {\bibinfo {author} {\bibfnamefont {A.}~\bibnamefont
  {Altland}}\ and\ \bibinfo {author} {\bibfnamefont {M.~R.}\ \bibnamefont
  {Zirnbauer}},\ }\bibfield  {title} {\bibinfo {title} {{Nonstandard symmetry
  classes in mesoscopic normal-superconducting hybrid structures}},\ }\href
  {https://doi.org/10.1103/PhysRevB.55.1142} {\bibfield  {journal} {\bibinfo
  {journal} {Phys. Rev. B}\ }\textbf {\bibinfo {volume} {55}},\ \bibinfo
  {pages} {1142} (\bibinfo {year} {1997})}\BibitemShut {NoStop}%
\bibitem [{\citenamefont {Polyakov}(1975)}]{Polyakov-75}%
  \BibitemOpen
  \bibfield  {author} {\bibinfo {author} {\bibfnamefont {A.~M.}\ \bibnamefont
  {Polyakov}},\ }\bibfield  {title} {\bibinfo {title} {{Interaction of
  goldstone particles in two dimensions. Applications to ferromagnets and
  massive Yang-Mills fields}},\ }\href
  {https://doi.org/10.1016/0370-2693(75)90161-6} {\bibfield  {journal}
  {\bibinfo  {journal} {Phys. Lett. B}\ }\textbf {\bibinfo {volume} {59}},\
  \bibinfo {pages} {79} (\bibinfo {year} {1975})}\BibitemShut {NoStop}%
\bibitem [{\citenamefont {Br\'ezin}\ and\ \citenamefont
  {Zinn-Justin}(1976{\natexlab{a}})}]{Brezin-76L}%
  \BibitemOpen
  \bibfield  {author} {\bibinfo {author} {\bibfnamefont {E.}~\bibnamefont
  {Br\'ezin}}\ and\ \bibinfo {author} {\bibfnamefont {J.}~\bibnamefont
  {Zinn-Justin}},\ }\bibfield  {title} {\bibinfo {title} {{Renormalization of
  the Nonlinear $\ensuremath{\sigma}$ Model in $2+\ensuremath{\epsilon}$
  Dimensions---Application to the Heisenberg Ferromagnets}},\ }\href
  {https://doi.org/10.1103/PhysRevLett.36.691} {\bibfield  {journal} {\bibinfo
  {journal} {Phys. Rev. Lett.}\ }\textbf {\bibinfo {volume} {36}},\ \bibinfo
  {pages} {691} (\bibinfo {year} {1976}{\natexlab{a}})}\BibitemShut {NoStop}%
\bibitem [{\citenamefont {Br\'ezin}\ and\ \citenamefont
  {Zinn-Justin}(1976{\natexlab{b}})}]{Brezin-76B}%
  \BibitemOpen
  \bibfield  {author} {\bibinfo {author} {\bibfnamefont {E.}~\bibnamefont
  {Br\'ezin}}\ and\ \bibinfo {author} {\bibfnamefont {J.}~\bibnamefont
  {Zinn-Justin}},\ }\bibfield  {title} {\bibinfo {title} {{Spontaneous
  breakdown of continuous symmetries near two dimensions}},\ }\href
  {https://doi.org/10.1103/PhysRevB.14.3110} {\bibfield  {journal} {\bibinfo
  {journal} {Phys. Rev. B}\ }\textbf {\bibinfo {volume} {14}},\ \bibinfo
  {pages} {3110} (\bibinfo {year} {1976}{\natexlab{b}})}\BibitemShut {NoStop}%
\bibitem [{\citenamefont {Wegner}(1979)}]{Wegner-79}%
  \BibitemOpen
  \bibfield  {author} {\bibinfo {author} {\bibfnamefont {F.}~\bibnamefont
  {Wegner}},\ }\bibfield  {title} {\bibinfo {title} {{The mobility edge
  problem: Continuous symmetry and a conjecture}},\ }\href
  {https://doi.org/10.1007/BF01319839} {\bibfield  {journal} {\bibinfo
  {journal} {Z. Physik B}\ }\textbf {\bibinfo {volume} {35}},\ \bibinfo {pages}
  {207} (\bibinfo {year} {1979})}\BibitemShut {NoStop}%
\bibitem [{\citenamefont {Sch\"afer}\ and\ \citenamefont
  {Wegner}(1980)}]{Schafer-80}%
  \BibitemOpen
  \bibfield  {author} {\bibinfo {author} {\bibfnamefont {L.}~\bibnamefont
  {Sch\"afer}}\ and\ \bibinfo {author} {\bibfnamefont {F.}~\bibnamefont
  {Wegner}},\ }\bibfield  {title} {\bibinfo {title} {{Disordered system with
  $n$ orbitals per site: Lagrange formulation, hyperbolic symmetry, and
  goldstone modes}},\ }\href {https://doi.org/10.1007/BF01598751} {\bibfield
  {journal} {\bibinfo  {journal} {Z. Physik B}\ }\textbf {\bibinfo {volume}
  {38}},\ \bibinfo {pages} {113} (\bibinfo {year} {1980})}\BibitemShut
  {NoStop}%
\bibitem [{\citenamefont {Gor'kov}\ \emph {et~al.}(1979)\citenamefont
  {Gor'kov}, \citenamefont {Larkin},\ and\ \citenamefont
  {Khmel'nitskii}}]{Gorkov-79}%
  \BibitemOpen
  \bibfield  {author} {\bibinfo {author} {\bibfnamefont {L.~P.}\ \bibnamefont
  {Gor'kov}}, \bibinfo {author} {\bibfnamefont {A.~I.}\ \bibnamefont
  {Larkin}},\ and\ \bibinfo {author} {\bibfnamefont {D.~E.}\ \bibnamefont
  {Khmel'nitskii}},\ }\bibfield  {title} {\bibinfo {title} {{Particle
  conductivity in a two-dimensional random potential}},\ }\href
  {http://jetpletters.ru/ps/0/article_20629.shtml} {\bibfield  {journal}
  {\bibinfo  {journal} {JETP Lett.}\ }\textbf {\bibinfo {volume} {30}},\
  \bibinfo {pages} {228} (\bibinfo {year} {1979})}\BibitemShut {NoStop}%
\bibitem [{\citenamefont {Hikami}\ \emph {et~al.}(1980)\citenamefont {Hikami},
  \citenamefont {Larkin},\ and\ \citenamefont {Nagaoka}}]{Hikami-80}%
  \BibitemOpen
  \bibfield  {author} {\bibinfo {author} {\bibfnamefont {S.}~\bibnamefont
  {Hikami}}, \bibinfo {author} {\bibfnamefont {A.~I.}\ \bibnamefont {Larkin}},\
  and\ \bibinfo {author} {\bibfnamefont {Y.}~\bibnamefont {Nagaoka}},\
  }\bibfield  {title} {\bibinfo {title} {{Spin-Orbit Interaction and
  Magnetoresistance in the Two Dimensional Random System}},\ }\href
  {https://doi.org/10.1143/PTP.63.707} {\bibfield  {journal} {\bibinfo
  {journal} {Prog. Theor. Phys.}\ }\textbf {\bibinfo {volume} {63}},\ \bibinfo
  {pages} {707} (\bibinfo {year} {1980})}\BibitemShut {NoStop}%
\bibitem [{\citenamefont {Efetov}\ \emph {et~al.}(1980)\citenamefont {Efetov},
  \citenamefont {Larkin},\ and\ \citenamefont {Khmel'nitzkii}}]{Efetov-80}%
  \BibitemOpen
  \bibfield  {author} {\bibinfo {author} {\bibfnamefont {K.~B.}\ \bibnamefont
  {Efetov}}, \bibinfo {author} {\bibfnamefont {A.~I.}\ \bibnamefont {Larkin}},\
  and\ \bibinfo {author} {\bibfnamefont {D.~E.}\ \bibnamefont
  {Khmel'nitzkii}},\ }\bibfield  {title} {\bibinfo {title} {{Interaction of
  diffusion modes in the theory of localization}},\ }\href
  {http://jetp.ras.ru/cgi-bin/e/index/e/52/3/p568?a=list} {\bibfield  {journal}
  {\bibinfo  {journal} {J. Exp. Theor. Phys.}\ }\textbf {\bibinfo {volume}
  {52}},\ \bibinfo {pages} {568} (\bibinfo {year} {1980})}\BibitemShut
  {NoStop}%
\bibitem [{\citenamefont {Altshuler}\ \emph {et~al.}(1980)\citenamefont
  {Altshuler}, \citenamefont {Khmel'nitzkii}, \citenamefont {Larkin},\ and\
  \citenamefont {Lee}}]{Altshuler-80}%
  \BibitemOpen
  \bibfield  {author} {\bibinfo {author} {\bibfnamefont {B.~L.}\ \bibnamefont
  {Altshuler}}, \bibinfo {author} {\bibfnamefont {D.}~\bibnamefont
  {Khmel'nitzkii}}, \bibinfo {author} {\bibfnamefont {A.~I.}\ \bibnamefont
  {Larkin}},\ and\ \bibinfo {author} {\bibfnamefont {P.~A.}\ \bibnamefont
  {Lee}},\ }\bibfield  {title} {\bibinfo {title} {{Magnetoresistance and Hall
  effect in a disordered two-dimensional electron gas}},\ }\href
  {https://doi.org/10.1103/PhysRevB.22.5142} {\bibfield  {journal} {\bibinfo
  {journal} {Phys. Rev. B}\ }\textbf {\bibinfo {volume} {22}},\ \bibinfo
  {pages} {5142} (\bibinfo {year} {1980})}\BibitemShut {NoStop}%
\bibitem [{\citenamefont {Dyson}(1953)}]{Dyson-53}%
  \BibitemOpen
  \bibfield  {author} {\bibinfo {author} {\bibfnamefont {F.~J.}\ \bibnamefont
  {Dyson}},\ }\bibfield  {title} {\bibinfo {title} {{The Dynamics of a
  Disordered Linear Chain}},\ }\href {https://doi.org/10.1103/PhysRev.92.1331}
  {\bibfield  {journal} {\bibinfo  {journal} {Phys. Rev.}\ }\textbf {\bibinfo
  {volume} {92}},\ \bibinfo {pages} {1331} (\bibinfo {year}
  {1953})}\BibitemShut {NoStop}%
\bibitem [{\citenamefont {Anderson}(1958)}]{Anderson-58}%
  \BibitemOpen
  \bibfield  {author} {\bibinfo {author} {\bibfnamefont {P.~W.}\ \bibnamefont
  {Anderson}},\ }\bibfield  {title} {\bibinfo {title} {{Absence of Diffusion in
  Certain Random Lattices}},\ }\href {https://doi.org/10.1103/PhysRev.109.1492}
  {\bibfield  {journal} {\bibinfo  {journal} {Phys. Rev.}\ }\textbf {\bibinfo
  {volume} {109}},\ \bibinfo {pages} {1492} (\bibinfo {year}
  {1958})}\BibitemShut {NoStop}%
\bibitem [{\citenamefont {Abrahams}\ \emph {et~al.}(1979)\citenamefont
  {Abrahams}, \citenamefont {Anderson}, \citenamefont {Licciardello},\ and\
  \citenamefont {Ramakrishnan}}]{Abrahams-79}%
  \BibitemOpen
  \bibfield  {author} {\bibinfo {author} {\bibfnamefont {E.}~\bibnamefont
  {Abrahams}}, \bibinfo {author} {\bibfnamefont {P.~W.}\ \bibnamefont
  {Anderson}}, \bibinfo {author} {\bibfnamefont {D.~C.}\ \bibnamefont
  {Licciardello}},\ and\ \bibinfo {author} {\bibfnamefont {T.~V.}\ \bibnamefont
  {Ramakrishnan}},\ }\bibfield  {title} {\bibinfo {title} {{Scaling Theory of
  Localization: Absence of Quantum Diffusion in Two Dimensions}},\ }\href
  {https://doi.org/10.1103/PhysRevLett.42.673} {\bibfield  {journal} {\bibinfo
  {journal} {Phys. Rev. Lett.}\ }\textbf {\bibinfo {volume} {42}},\ \bibinfo
  {pages} {673} (\bibinfo {year} {1979})}\BibitemShut {NoStop}%
\bibitem [{\citenamefont {Lee}\ and\ \citenamefont
  {Ramakrishnan}(1985)}]{Lee-review}%
  \BibitemOpen
  \bibfield  {author} {\bibinfo {author} {\bibfnamefont {P.~A.}\ \bibnamefont
  {Lee}}\ and\ \bibinfo {author} {\bibfnamefont {T.~V.}\ \bibnamefont
  {Ramakrishnan}},\ }\bibfield  {title} {\bibinfo {title} {{Disordered
  electronic systems}},\ }\href {https://doi.org/10.1103/RevModPhys.57.287}
  {\bibfield  {journal} {\bibinfo  {journal} {Rev. Mod. Phys.}\ }\textbf
  {\bibinfo {volume} {57}},\ \bibinfo {pages} {287} (\bibinfo {year}
  {1985})}\BibitemShut {NoStop}%
\bibitem [{\citenamefont {Evers}\ and\ \citenamefont
  {Mirlin}(2008)}]{Evers-review}%
  \BibitemOpen
  \bibfield  {author} {\bibinfo {author} {\bibfnamefont {F.}~\bibnamefont
  {Evers}}\ and\ \bibinfo {author} {\bibfnamefont {A.~D.}\ \bibnamefont
  {Mirlin}},\ }\bibfield  {title} {\bibinfo {title} {{Anderson transitions}},\
  }\href {https://doi.org/10.1103/RevModPhys.80.1355} {\bibfield  {journal}
  {\bibinfo  {journal} {Rev. Mod. Phys.}\ }\textbf {\bibinfo {volume} {80}},\
  \bibinfo {pages} {1355} (\bibinfo {year} {2008})}\BibitemShut {NoStop}%
\bibitem [{\citenamefont {Schnyder}\ \emph {et~al.}(2008)\citenamefont
  {Schnyder}, \citenamefont {Ryu}, \citenamefont {Furusaki},\ and\
  \citenamefont {Ludwig}}]{Schnyder-08}%
  \BibitemOpen
  \bibfield  {author} {\bibinfo {author} {\bibfnamefont {A.~P.}\ \bibnamefont
  {Schnyder}}, \bibinfo {author} {\bibfnamefont {S.}~\bibnamefont {Ryu}},
  \bibinfo {author} {\bibfnamefont {A.}~\bibnamefont {Furusaki}},\ and\
  \bibinfo {author} {\bibfnamefont {A.~W.~W.}\ \bibnamefont {Ludwig}},\
  }\bibfield  {title} {\bibinfo {title} {{Classification of topological
  insulators and superconductors in three spatial dimensions}},\ }\href
  {https://doi.org/10.1103/PhysRevB.78.195125} {\bibfield  {journal} {\bibinfo
  {journal} {Phys. Rev. B}\ }\textbf {\bibinfo {volume} {78}},\ \bibinfo
  {pages} {195125} (\bibinfo {year} {2008})}\BibitemShut {NoStop}%
\bibitem [{\citenamefont {Kitaev}(2009)}]{Kitaev-09}%
  \BibitemOpen
  \bibfield  {author} {\bibinfo {author} {\bibfnamefont {A.}~\bibnamefont
  {Kitaev}},\ }\bibfield  {title} {\bibinfo {title} {{Periodic table for
  topological insulators and superconductors}},\ }\href
  {https://doi.org/10.1063/1.3149495} {\bibfield  {journal} {\bibinfo
  {journal} {AIP Conf. Proc.}\ }\textbf {\bibinfo {volume} {1134}},\ \bibinfo
  {pages} {22} (\bibinfo {year} {2009})}\BibitemShut {NoStop}%
\bibitem [{\citenamefont {Ryu}\ \emph {et~al.}(2010)\citenamefont {Ryu},
  \citenamefont {Schnyder}, \citenamefont {Furusaki},\ and\ \citenamefont
  {Ludwig}}]{Ryu-10}%
  \BibitemOpen
  \bibfield  {author} {\bibinfo {author} {\bibfnamefont {S.}~\bibnamefont
  {Ryu}}, \bibinfo {author} {\bibfnamefont {A.~P.}\ \bibnamefont {Schnyder}},
  \bibinfo {author} {\bibfnamefont {A.}~\bibnamefont {Furusaki}},\ and\
  \bibinfo {author} {\bibfnamefont {A.~W.~W.}\ \bibnamefont {Ludwig}},\
  }\bibfield  {title} {\bibinfo {title} {{Topological insulators and
  superconductors: tenfold way and dimensional hierarchy}},\ }\href
  {https://doi.org/10.1088/1367-2630/12/6/065010} {\bibfield  {journal}
  {\bibinfo  {journal} {New J. Phys.}\ }\textbf {\bibinfo {volume} {12}},\
  \bibinfo {pages} {065010} (\bibinfo {year} {2010})}\BibitemShut {NoStop}%
\bibitem [{\citenamefont {Chiu}\ \emph {et~al.}(2016)\citenamefont {Chiu},
  \citenamefont {Teo}, \citenamefont {Schnyder},\ and\ \citenamefont
  {Ryu}}]{CTSR-review}%
  \BibitemOpen
  \bibfield  {author} {\bibinfo {author} {\bibfnamefont {C.-K.}\ \bibnamefont
  {Chiu}}, \bibinfo {author} {\bibfnamefont {J.~C.~Y.}\ \bibnamefont {Teo}},
  \bibinfo {author} {\bibfnamefont {A.~P.}\ \bibnamefont {Schnyder}},\ and\
  \bibinfo {author} {\bibfnamefont {S.}~\bibnamefont {Ryu}},\ }\bibfield
  {title} {\bibinfo {title} {{Classification of topological quantum matter with
  symmetries}},\ }\href {https://doi.org/10.1103/RevModPhys.88.035005}
  {\bibfield  {journal} {\bibinfo  {journal} {Rev. Mod. Phys.}\ }\textbf
  {\bibinfo {volume} {88}},\ \bibinfo {pages} {035005} (\bibinfo {year}
  {2016})}\BibitemShut {NoStop}%
\bibitem [{\citenamefont {Haldane}(1983{\natexlab{a}})}]{Haldane-83PLA}%
  \BibitemOpen
  \bibfield  {author} {\bibinfo {author} {\bibfnamefont {F.~D.~M.}\
  \bibnamefont {Haldane}},\ }\bibfield  {title} {\bibinfo {title} {{Continuum
  dynamics of the 1-D Heisenberg antiferromagnet: Identification with the O(3)
  nonlinear sigma model}},\ }\href
  {https://doi.org/10.1103/RevModPhys.88.035005} {\bibfield  {journal}
  {\bibinfo  {journal} {Phys. Lett. A}\ }\textbf {\bibinfo {volume} {93}},\
  \bibinfo {pages} {464} (\bibinfo {year} {1983}{\natexlab{a}})}\BibitemShut
  {NoStop}%
\bibitem [{\citenamefont {Haldane}(1983{\natexlab{b}})}]{Haldane-83PRL}%
  \BibitemOpen
  \bibfield  {author} {\bibinfo {author} {\bibfnamefont {F.~D.~M.}\
  \bibnamefont {Haldane}},\ }\bibfield  {title} {\bibinfo {title} {{Nonlinear
  Field Theory of Large-Spin Heisenberg Antiferromagnets: Semiclassically
  Quantized Solitons of the One-Dimensional Easy-Axis N\'eel State}},\ }\href
  {https://doi.org/10.1103/PhysRevLett.50.1153} {\bibfield  {journal} {\bibinfo
   {journal} {Phys. Rev. Lett.}\ }\textbf {\bibinfo {volume} {50}},\ \bibinfo
  {pages} {1153} (\bibinfo {year} {1983}{\natexlab{b}})}\BibitemShut {NoStop}%
\bibitem [{\citenamefont {Affleck}(1989)}]{Affleck-89}%
  \BibitemOpen
  \bibfield  {author} {\bibinfo {author} {\bibfnamefont {I.}~\bibnamefont
  {Affleck}},\ }\bibfield  {title} {\bibinfo {title} {{Quantum spin chains and
  the Haldane gap}},\ }\href {https://doi.org/10.1088/0953-8984/1/19/001}
  {\bibfield  {journal} {\bibinfo  {journal} {J. Phys.: Condens. Matter}\
  }\textbf {\bibinfo {volume} {1}},\ \bibinfo {pages} {3047} (\bibinfo {year}
  {1989})}\BibitemShut {NoStop}%
\bibitem [{\citenamefont {Tasaki}(2020)}]{Tasaki-textbook}%
  \BibitemOpen
  \bibfield  {author} {\bibinfo {author} {\bibfnamefont {H.}~\bibnamefont
  {Tasaki}},\ }\href {https://doi.org/10.1007/978-3-030-41265-4} {\emph
  {\bibinfo {title} {{Physics and Mathematics of Quantum Many-Body Systems}}}}\
  (\bibinfo  {publisher} {Springer},\ \bibinfo {address} {Cham},\ \bibinfo
  {year} {2020})\BibitemShut {NoStop}%
\bibitem [{\citenamefont {Jian}\ \emph {et~al.}(2022)\citenamefont {Jian},
  \citenamefont {Bauer}, \citenamefont {Keselman},\ and\ \citenamefont
  {Ludwig}}]{Jian-22}%
  \BibitemOpen
  \bibfield  {author} {\bibinfo {author} {\bibfnamefont {C.-M.}\ \bibnamefont
  {Jian}}, \bibinfo {author} {\bibfnamefont {B.}~\bibnamefont {Bauer}},
  \bibinfo {author} {\bibfnamefont {A.}~\bibnamefont {Keselman}},\ and\
  \bibinfo {author} {\bibfnamefont {A.~W.~W.}\ \bibnamefont {Ludwig}},\
  }\bibfield  {title} {\bibinfo {title} {{Criticality and entanglement in
  nonunitary quantum circuits and tensor networks of noninteracting
  fermions}},\ }\href {https://doi.org/10.1103/PhysRevB.106.134206} {\bibfield
  {journal} {\bibinfo  {journal} {Phys. Rev. B}\ }\textbf {\bibinfo {volume}
  {106}},\ \bibinfo {pages} {134206} (\bibinfo {year} {2022})}\BibitemShut
  {NoStop}%
\bibitem [{\citenamefont {Jian}\ \emph {et~al.}()\citenamefont {Jian},
  \citenamefont {Shapourian}, \citenamefont {Bauer},\ and\ \citenamefont
  {Ludwig}}]{Jian-23}%
  \BibitemOpen
  \bibfield  {author} {\bibinfo {author} {\bibfnamefont {C.-M.}\ \bibnamefont
  {Jian}}, \bibinfo {author} {\bibfnamefont {H.}~\bibnamefont {Shapourian}},
  \bibinfo {author} {\bibfnamefont {B.}~\bibnamefont {Bauer}},\ and\ \bibinfo
  {author} {\bibfnamefont {A.~W.~W.}\ \bibnamefont {Ludwig}},\ }\bibfield
  {title} {\bibinfo {title} {{Measurement-induced entanglement transitions in
  quantum circuits of non-interacting fermions: Born-rule versus forced
  measurements}},\ }\Eprint {https://arxiv.org/abs/2302.09094}
  {arXiv:2302.09094} \BibitemShut {NoStop}%
\bibitem [{\citenamefont {Yang}\ \emph {et~al.}(2023)\citenamefont {Yang},
  \citenamefont {Zuo},\ and\ \citenamefont {Liu}}]{Yang-23}%
  \BibitemOpen
  \bibfield  {author} {\bibinfo {author} {\bibfnamefont {Q.}~\bibnamefont
  {Yang}}, \bibinfo {author} {\bibfnamefont {Y.}~\bibnamefont {Zuo}},\ and\
  \bibinfo {author} {\bibfnamefont {D.~E.}\ \bibnamefont {Liu}},\ }\bibfield
  {title} {\bibinfo {title} {{Keldysh nonlinear sigma model for a free-fermion
  gas under continuous measurements}},\ }\href
  {https://doi.org/10.1103/PhysRevResearch.5.033174} {\bibfield  {journal}
  {\bibinfo  {journal} {Phys. Rev. Research}\ }\textbf {\bibinfo {volume}
  {5}},\ \bibinfo {pages} {033174} (\bibinfo {year} {2023})}\BibitemShut
  {NoStop}%
\bibitem [{\citenamefont {Fava}\ \emph {et~al.}(2023)\citenamefont {Fava},
  \citenamefont {Piroli}, \citenamefont {Swann}, \citenamefont {Bernard},\ and\
  \citenamefont {Nahum}}]{Fava-23}%
  \BibitemOpen
  \bibfield  {author} {\bibinfo {author} {\bibfnamefont {M.}~\bibnamefont
  {Fava}}, \bibinfo {author} {\bibfnamefont {L.}~\bibnamefont {Piroli}},
  \bibinfo {author} {\bibfnamefont {T.}~\bibnamefont {Swann}}, \bibinfo
  {author} {\bibfnamefont {D.}~\bibnamefont {Bernard}},\ and\ \bibinfo {author}
  {\bibfnamefont {A.}~\bibnamefont {Nahum}},\ }\bibfield  {title} {\bibinfo
  {title} {{Nonlinear Sigma Models for Monitored Dynamics of Free Fermions}},\
  }\href {https://doi.org/10.1103/PhysRevX.13.041045} {\bibfield  {journal}
  {\bibinfo  {journal} {Phys. Rev. X}\ }\textbf {\bibinfo {volume} {13}},\
  \bibinfo {pages} {041045} (\bibinfo {year} {2023})}\BibitemShut {NoStop}%
\bibitem [{\citenamefont {Poboiko}\ \emph {et~al.}(2023)\citenamefont
  {Poboiko}, \citenamefont {P\"opperl}, \citenamefont {Gornyi},\ and\
  \citenamefont {Mirlin}}]{Poboiko-23}%
  \BibitemOpen
  \bibfield  {author} {\bibinfo {author} {\bibfnamefont {I.}~\bibnamefont
  {Poboiko}}, \bibinfo {author} {\bibfnamefont {P.}~\bibnamefont {P\"opperl}},
  \bibinfo {author} {\bibfnamefont {I.~V.}\ \bibnamefont {Gornyi}},\ and\
  \bibinfo {author} {\bibfnamefont {A.~D.}\ \bibnamefont {Mirlin}},\ }\bibfield
   {title} {\bibinfo {title} {{Theory of Free Fermions under Random Projective
  Measurements}},\ }\href {https://doi.org/10.1103/PhysRevX.13.041046}
  {\bibfield  {journal} {\bibinfo  {journal} {Phys. Rev. X}\ }\textbf {\bibinfo
  {volume} {13}},\ \bibinfo {pages} {041046} (\bibinfo {year}
  {2023})}\BibitemShut {NoStop}%
\bibitem [{\citenamefont {Konotop}\ \emph {et~al.}(2016)\citenamefont
  {Konotop}, \citenamefont {Yang},\ and\ \citenamefont
  {Zezyulin}}]{Konotop-review}%
  \BibitemOpen
  \bibfield  {author} {\bibinfo {author} {\bibfnamefont {V.~V.}\ \bibnamefont
  {Konotop}}, \bibinfo {author} {\bibfnamefont {J.}~\bibnamefont {Yang}},\ and\
  \bibinfo {author} {\bibfnamefont {D.~A.}\ \bibnamefont {Zezyulin}},\
  }\bibfield  {title} {\bibinfo {title} {{Nonlinear waves in
  $\mathcal{PT}$-symmetric systems}},\ }\href
  {https://doi.org/10.1103/RevModPhys.88.035002} {\bibfield  {journal}
  {\bibinfo  {journal} {Rev. Mod. Phys.}\ }\textbf {\bibinfo {volume} {88}},\
  \bibinfo {pages} {035002} (\bibinfo {year} {2016})}\BibitemShut {NoStop}%
\bibitem [{\citenamefont {El-Ganainy}\ \emph {et~al.}(2018)\citenamefont
  {El-Ganainy}, \citenamefont {Makris}, \citenamefont {Khajavikhan},
  \citenamefont {Musslimani}, \citenamefont {Rotter},\ and\ \citenamefont
  {Christodoulides}}]{Christodoulides-review}%
  \BibitemOpen
  \bibfield  {author} {\bibinfo {author} {\bibfnamefont {R.}~\bibnamefont
  {El-Ganainy}}, \bibinfo {author} {\bibfnamefont {K.~G.}\ \bibnamefont
  {Makris}}, \bibinfo {author} {\bibfnamefont {M.}~\bibnamefont {Khajavikhan}},
  \bibinfo {author} {\bibfnamefont {Z.~H.}\ \bibnamefont {Musslimani}},
  \bibinfo {author} {\bibfnamefont {S.}~\bibnamefont {Rotter}},\ and\ \bibinfo
  {author} {\bibfnamefont {D.~N.}\ \bibnamefont {Christodoulides}},\ }\bibfield
   {title} {\bibinfo {title} {{Non-Hermitian physics and PT symmetry}},\ }\href
  {https://doi.org/10.1038/nphys4323} {\bibfield  {journal} {\bibinfo
  {journal} {Nat. Phys.}\ }\textbf {\bibinfo {volume} {14}},\ \bibinfo {pages}
  {11} (\bibinfo {year} {2018})}\BibitemShut {NoStop}%
\bibitem [{\citenamefont {Bender}\ and\ \citenamefont
  {Boettcher}(1998)}]{Bender-02}%
  \BibitemOpen
  \bibfield  {author} {\bibinfo {author} {\bibfnamefont {C.~M.}\ \bibnamefont
  {Bender}}\ and\ \bibinfo {author} {\bibfnamefont {S.}~\bibnamefont
  {Boettcher}},\ }\bibfield  {title} {\bibinfo {title} {{Real Spectra in
  Non-Hermitian Hamiltonians Having $\mathcal{P}\mathcal{T}$ Symmetry}},\
  }\href {https://doi.org/10.1103/PhysRevLett.80.5243} {\bibfield  {journal}
  {\bibinfo  {journal} {Phys. Rev. Lett.}\ }\textbf {\bibinfo {volume} {80}},\
  \bibinfo {pages} {5243} (\bibinfo {year} {1998})}\BibitemShut {NoStop}%
\bibitem [{\citenamefont {Bender}(2007)}]{Bender-review}%
  \BibitemOpen
  \bibfield  {author} {\bibinfo {author} {\bibfnamefont {C.~M.}\ \bibnamefont
  {Bender}},\ }\bibfield  {title} {\bibinfo {title} {{Making sense of
  non-Hermitian Hamiltonians}},\ }\href
  {https://doi.org/10.1088/0034-4885/70/6/R03} {\bibfield  {journal} {\bibinfo
  {journal} {Rep. Prog. Phys.}\ }\textbf {\bibinfo {volume} {70}},\ \bibinfo
  {pages} {947} (\bibinfo {year} {2007})}\BibitemShut {NoStop}%
\bibitem [{\citenamefont {Yang}\ and\ \citenamefont {Lee}(1952)}]{Yang-52}%
  \BibitemOpen
  \bibfield  {author} {\bibinfo {author} {\bibfnamefont {C.~N.}\ \bibnamefont
  {Yang}}\ and\ \bibinfo {author} {\bibfnamefont {T.~D.}\ \bibnamefont {Lee}},\
  }\bibfield  {title} {\bibinfo {title} {{Statistical Theory of Equations of
  State and Phase Transitions. I. Theory of Condensation}},\ }\href
  {https://doi.org/10.1103/PhysRev.87.404} {\bibfield  {journal} {\bibinfo
  {journal} {Phys. Rev.}\ }\textbf {\bibinfo {volume} {87}},\ \bibinfo {pages}
  {404} (\bibinfo {year} {1952})}\BibitemShut {NoStop}%
\bibitem [{\citenamefont {Lee}\ and\ \citenamefont {Yang}(1952)}]{Lee-52}%
  \BibitemOpen
  \bibfield  {author} {\bibinfo {author} {\bibfnamefont {T.~D.}\ \bibnamefont
  {Lee}}\ and\ \bibinfo {author} {\bibfnamefont {C.~N.}\ \bibnamefont {Yang}},\
  }\bibfield  {title} {\bibinfo {title} {{Statistical Theory of Equations of
  State and Phase Transitions. II. Lattice Gas and Ising Model}},\ }\href
  {https://doi.org/10.1103/PhysRev.87.410} {\bibfield  {journal} {\bibinfo
  {journal} {Phys. Rev.}\ }\textbf {\bibinfo {volume} {87}},\ \bibinfo {pages}
  {410} (\bibinfo {year} {1952})}\BibitemShut {NoStop}%
\bibitem [{\citenamefont {Fisher}(1978)}]{Fisher-78}%
  \BibitemOpen
  \bibfield  {author} {\bibinfo {author} {\bibfnamefont {M.~E.}\ \bibnamefont
  {Fisher}},\ }\bibfield  {title} {\bibinfo {title} {{Yang-Lee Edge Singularity
  and ${\ensuremath{\phi}}^{3}$ Field Theory}},\ }\href
  {https://doi.org/10.1103/PhysRevLett.40.1610} {\bibfield  {journal} {\bibinfo
   {journal} {Phys. Rev. Lett.}\ }\textbf {\bibinfo {volume} {40}},\ \bibinfo
  {pages} {1610} (\bibinfo {year} {1978})}\BibitemShut {NoStop}%
\bibitem [{\citenamefont {Cardy}()}]{Cardy-23}%
  \BibitemOpen
  \bibfield  {author} {\bibinfo {author} {\bibfnamefont {J.}~\bibnamefont
  {Cardy}},\ }\bibfield  {title} {\bibinfo {title} {{The Yang-Lee Edge
  Singularity and Related Problems}},\ }\Eprint
  {https://arxiv.org/abs/2305.13288} {arXiv:2305.13288} \BibitemShut {NoStop}%
\bibitem [{\citenamefont {Cardy}(1985)}]{Cardy-85}%
  \BibitemOpen
  \bibfield  {author} {\bibinfo {author} {\bibfnamefont {J.~L.}\ \bibnamefont
  {Cardy}},\ }\bibfield  {title} {\bibinfo {title} {{Conformal Invariance and
  the Yang-Lee Edge Singularity in Two Dimensions}},\ }\href
  {https://doi.org/10.1103/PhysRevLett.54.1354} {\bibfield  {journal} {\bibinfo
   {journal} {Phys. Rev. Lett.}\ }\textbf {\bibinfo {volume} {54}},\ \bibinfo
  {pages} {1354} (\bibinfo {year} {1985})}\BibitemShut {NoStop}%
\bibitem [{\citenamefont {Bianchini}\ \emph {et~al.}(2015)\citenamefont
  {Bianchini}, \citenamefont {Castro-Alvaredo}, \citenamefont {Doyon},
  \citenamefont {Levi},\ and\ \citenamefont {Ravanini}}]{Bianchini-15}%
  \BibitemOpen
  \bibfield  {author} {\bibinfo {author} {\bibfnamefont {D.}~\bibnamefont
  {Bianchini}}, \bibinfo {author} {\bibfnamefont {O.}~\bibnamefont
  {Castro-Alvaredo}}, \bibinfo {author} {\bibfnamefont {B.}~\bibnamefont
  {Doyon}}, \bibinfo {author} {\bibfnamefont {E.}~\bibnamefont {Levi}},\ and\
  \bibinfo {author} {\bibfnamefont {F.}~\bibnamefont {Ravanini}},\ }\bibfield
  {title} {\bibinfo {title} {{Entanglement entropy of non-unitary conformal
  field theory}},\ }\href {https://doi.org/10.1088/1751-8113/48/4/04FT01}
  {\bibfield  {journal} {\bibinfo  {journal} {J. Phys. A}\ }\textbf {\bibinfo
  {volume} {48}},\ \bibinfo {pages} {04FT01} (\bibinfo {year}
  {2015})}\BibitemShut {NoStop}%
\bibitem [{\citenamefont {Couvreur}\ \emph {et~al.}(2017)\citenamefont
  {Couvreur}, \citenamefont {Jacobsen},\ and\ \citenamefont
  {Saleur}}]{Couvreur-17}%
  \BibitemOpen
  \bibfield  {author} {\bibinfo {author} {\bibfnamefont {R.}~\bibnamefont
  {Couvreur}}, \bibinfo {author} {\bibfnamefont {J.~L.}\ \bibnamefont
  {Jacobsen}},\ and\ \bibinfo {author} {\bibfnamefont {H.}~\bibnamefont
  {Saleur}},\ }\bibfield  {title} {\bibinfo {title} {{Entanglement in
  Nonunitary Quantum Critical Spin Chains}},\ }\href
  {https://doi.org/10.1103/PhysRevLett.119.040601} {\bibfield  {journal}
  {\bibinfo  {journal} {Phys. Rev. Lett.}\ }\textbf {\bibinfo {volume} {119}},\
  \bibinfo {pages} {040601} (\bibinfo {year} {2017})}\BibitemShut {NoStop}%
\bibitem [{\citenamefont {Chang}\ \emph {et~al.}(2020)\citenamefont {Chang},
  \citenamefont {You}, \citenamefont {Wen},\ and\ \citenamefont
  {Ryu}}]{Chang-20}%
  \BibitemOpen
  \bibfield  {author} {\bibinfo {author} {\bibfnamefont {P.-Y.}\ \bibnamefont
  {Chang}}, \bibinfo {author} {\bibfnamefont {J.-S.}\ \bibnamefont {You}},
  \bibinfo {author} {\bibfnamefont {X.}~\bibnamefont {Wen}},\ and\ \bibinfo
  {author} {\bibfnamefont {S.}~\bibnamefont {Ryu}},\ }\bibfield  {title}
  {\bibinfo {title} {{Entanglement spectrum and entropy in topological
  non-Hermitian systems and nonunitary conformal field theory}},\ }\href
  {https://doi.org/10.1103/PhysRevResearch.2.033069} {\bibfield  {journal}
  {\bibinfo  {journal} {Phys. Rev. Research}\ }\textbf {\bibinfo {volume}
  {2}},\ \bibinfo {pages} {033069} (\bibinfo {year} {2020})}\BibitemShut
  {NoStop}%
\bibitem [{\citenamefont {Lee}(2022)}]{Lee-22}%
  \BibitemOpen
  \bibfield  {author} {\bibinfo {author} {\bibfnamefont {C.~H.}\ \bibnamefont
  {Lee}},\ }\bibfield  {title} {\bibinfo {title} {{Exceptional Bound States and
  Negative Entanglement Entropy}},\ }\href
  {https://doi.org/10.1103/PhysRevLett.128.010402} {\bibfield  {journal}
  {\bibinfo  {journal} {Phys. Rev. Lett.}\ }\textbf {\bibinfo {volume} {128}},\
  \bibinfo {pages} {010402} (\bibinfo {year} {2022})}\BibitemShut {NoStop}%
\bibitem [{\citenamefont {Tu}\ \emph {et~al.}(2022)\citenamefont {Tu},
  \citenamefont {Tzeng},\ and\ \citenamefont {Chang}}]{Tu-22}%
  \BibitemOpen
  \bibfield  {author} {\bibinfo {author} {\bibfnamefont {Y.-T.}\ \bibnamefont
  {Tu}}, \bibinfo {author} {\bibfnamefont {Y.-C.}\ \bibnamefont {Tzeng}},\ and\
  \bibinfo {author} {\bibfnamefont {P.-Y.}\ \bibnamefont {Chang}},\ }\bibfield
  {title} {\bibinfo {title} {{R\'enyi entropies and negative central charges in
  non-Hermitian quantum systems}},\ }\href
  {https://doi.org/10.21468/SciPostPhys.12.6.194} {\bibfield  {journal}
  {\bibinfo  {journal} {SciPost Phys.}\ }\textbf {\bibinfo {volume} {12}},\
  \bibinfo {pages} {194} (\bibinfo {year} {2022})}\BibitemShut {NoStop}%
\bibitem [{\citenamefont {Ryu}\ and\ \citenamefont {Yoon}(2023)}]{Ryu-23}%
  \BibitemOpen
  \bibfield  {author} {\bibinfo {author} {\bibfnamefont {S.}~\bibnamefont
  {Ryu}}\ and\ \bibinfo {author} {\bibfnamefont {J.}~\bibnamefont {Yoon}},\
  }\bibfield  {title} {\bibinfo {title} {{Unitarity of Symplectic Fermions in
  $\ensuremath{\alpha}$ Vacua with Negative Central Charge}},\ }\href
  {https://doi.org/10.1103/PhysRevLett.130.241602} {\bibfield  {journal}
  {\bibinfo  {journal} {Phys. Rev. Lett.}\ }\textbf {\bibinfo {volume} {130}},\
  \bibinfo {pages} {241602} (\bibinfo {year} {2023})}\BibitemShut {NoStop}%
\bibitem [{\citenamefont {Hsieh}\ and\ \citenamefont {Chang}(2023)}]{Hsieh-23}%
  \BibitemOpen
  \bibfield  {author} {\bibinfo {author} {\bibfnamefont {C.-T.}\ \bibnamefont
  {Hsieh}}\ and\ \bibinfo {author} {\bibfnamefont {P.-Y.}\ \bibnamefont
  {Chang}},\ }\bibfield  {title} {\bibinfo {title} {{Relating non-Hermitian and
  Hermitian quantum systems at criticality}},\ }\href
  {https://doi.org/10.21468/SciPostPhysCore.6.3.062} {\bibfield  {journal}
  {\bibinfo  {journal} {SciPost Phys. Core}\ }\textbf {\bibinfo {volume} {6}},\
  \bibinfo {pages} {062} (\bibinfo {year} {2023})}\BibitemShut {NoStop}%
\bibitem [{\citenamefont {Fossati}\ \emph {et~al.}(2023)\citenamefont
  {Fossati}, \citenamefont {Ares},\ and\ \citenamefont
  {Calabrese}}]{Fossati-23}%
  \BibitemOpen
  \bibfield  {author} {\bibinfo {author} {\bibfnamefont {M.}~\bibnamefont
  {Fossati}}, \bibinfo {author} {\bibfnamefont {F.}~\bibnamefont {Ares}},\ and\
  \bibinfo {author} {\bibfnamefont {P.}~\bibnamefont {Calabrese}},\ }\bibfield
  {title} {\bibinfo {title} {{Symmetry-resolved entanglement in critical
  non-Hermitian systems}},\ }\href
  {https://doi.org/10.1103/PhysRevB.107.205153} {\bibfield  {journal} {\bibinfo
   {journal} {Phys. Rev. B}\ }\textbf {\bibinfo {volume} {107}},\ \bibinfo
  {pages} {205153} (\bibinfo {year} {2023})}\BibitemShut {NoStop}%
\bibitem [{\citenamefont {Rottoli}\ \emph {et~al.}(2024)\citenamefont
  {Rottoli}, \citenamefont {Fossati},\ and\ \citenamefont
  {Calabrese}}]{Rottoli-24}%
  \BibitemOpen
  \bibfield  {author} {\bibinfo {author} {\bibfnamefont {F.}~\bibnamefont
  {Rottoli}}, \bibinfo {author} {\bibfnamefont {M.}~\bibnamefont {Fossati}},\
  and\ \bibinfo {author} {\bibfnamefont {P.}~\bibnamefont {Calabrese}},\
  }\bibfield  {title} {\bibinfo {title} {{Entanglement Hamiltonian in the
  non-Hermitian SSH model}},\ }\href {https://doi.org/10.1088/1742-5468/ad4860}
  {\bibfield  {journal} {\bibinfo  {journal} {J. Stat. Mech.}\ }\textbf
  {\bibinfo {volume} {2024}},\ \bibinfo {pages} {063102} (\bibinfo {year}
  {2024})}\BibitemShut {NoStop}%
\bibitem [{\citenamefont {Xue}\ and\ \citenamefont {Lee}(2025)}]{Xue-25}%
  \BibitemOpen
  \bibfield  {author} {\bibinfo {author} {\bibfnamefont {W.-T.}\ \bibnamefont
  {Xue}}\ and\ \bibinfo {author} {\bibfnamefont {C.~H.}\ \bibnamefont {Lee}},\
  }\bibfield  {title} {\bibinfo {title} {{Topologically Protected Negative
  Entanglement}},\ }\href {https://doi.org/10.1002/advs.202513868} {\bibfield
  {journal} {\bibinfo  {journal} {Adv. Sci.}\ }\textbf {\bibinfo {volume}
  {2025}},\ \bibinfo {pages} {e13868} (\bibinfo {year} {2025})}\BibitemShut
  {NoStop}%
\bibitem [{\citenamefont {Li}\ \emph {et~al.}(2026)\citenamefont {Li},
  \citenamefont {Chou}, \citenamefont {Wen},\ and\ \citenamefont
  {Chang}}]{Li-25}%
  \BibitemOpen
  \bibfield  {author} {\bibinfo {author} {\bibfnamefont {H.-H.}\ \bibnamefont
  {Li}}, \bibinfo {author} {\bibfnamefont {K.-H.}\ \bibnamefont {Chou}},
  \bibinfo {author} {\bibfnamefont {X.}~\bibnamefont {Wen}},\ and\ \bibinfo
  {author} {\bibfnamefont {P.-Y.}\ \bibnamefont {Chang}},\ }\bibfield  {title}
  {\bibinfo {title} {{Impurity-induced nonunitary criticality}},\ }\href
  {https://doi.org/10.1103/m7lb-vkbf} {\bibfield  {journal} {\bibinfo
  {journal} {Phys. Rev. B}\ }\textbf {\bibinfo {volume} {113}},\ \bibinfo
  {pages} {035130} (\bibinfo {year} {2026})}\BibitemShut {NoStop}%
\bibitem [{\citenamefont {Fan}\ \emph {et~al.}()\citenamefont {Fan},
  \citenamefont {Dong},\ and\ \citenamefont {Vishwanath}}]{Fan-25}%
  \BibitemOpen
  \bibfield  {author} {\bibinfo {author} {\bibfnamefont {R.}~\bibnamefont
  {Fan}}, \bibinfo {author} {\bibfnamefont {J.}~\bibnamefont {Dong}},\ and\
  \bibinfo {author} {\bibfnamefont {A.}~\bibnamefont {Vishwanath}},\ }\bibfield
   {title} {\bibinfo {title} {{Simulating the non-unitary Yang-Lee conformal
  field theory on the fuzzy sphere}},\ }\Eprint
  {https://arxiv.org/abs/2505.06342} {arXiv:2505.06342} \BibitemShut {NoStop}%
\bibitem [{\citenamefont {Arguello~Cruz}\ \emph {et~al.}(2026)\citenamefont
  {Arguello~Cruz}, \citenamefont {Klebanov}, \citenamefont {Tarnopolsky},\ and\
  \citenamefont {Xin}}]{Cruz-25}%
  \BibitemOpen
  \bibfield  {author} {\bibinfo {author} {\bibfnamefont {E.}~\bibnamefont
  {Arguello~Cruz}}, \bibinfo {author} {\bibfnamefont {I.~R.}\ \bibnamefont
  {Klebanov}}, \bibinfo {author} {\bibfnamefont {G.}~\bibnamefont
  {Tarnopolsky}},\ and\ \bibinfo {author} {\bibfnamefont {Y.}~\bibnamefont
  {Xin}},\ }\bibfield  {title} {\bibinfo {title} {{Yang-Lee Quantum Criticality
  in Various Dimensions}},\ }\href {https://doi.org/10.1103/w4qg-2xwn}
  {\bibfield  {journal} {\bibinfo  {journal} {Phys. Rev. X}\ }\textbf {\bibinfo
  {volume} {16}},\ \bibinfo {pages} {011022} (\bibinfo {year}
  {2026})}\BibitemShut {NoStop}%
\bibitem [{\citenamefont {Miró}\ and\ \citenamefont
  {Delouche}(2025)}]{Miro-25}%
  \BibitemOpen
  \bibfield  {author} {\bibinfo {author} {\bibfnamefont {J.~E.}\ \bibnamefont
  {Miró}}\ and\ \bibinfo {author} {\bibfnamefont {O.}~\bibnamefont
  {Delouche}},\ }\bibfield  {title} {\bibinfo {title} {{Flowing from the Ising
  model on the fuzzy sphere to the 3D Lee-Yang CFT}},\ }\href
  {https://doi.org/10.1007/JHEP10(2025)037} {\bibfield  {journal} {\bibinfo
  {journal} {J. High Energ. Phys.}\ }\textbf {\bibinfo {volume} {2025}}\bibinfo
   {number} { (10)},\ \bibinfo {pages} {37}}\BibitemShut {NoStop}%
\bibitem [{\citenamefont {Chou}\ \emph {et~al.}()\citenamefont {Chou},
  \citenamefont {Yu},\ and\ \citenamefont {Chang}}]{Chou-25}%
  \BibitemOpen
\bibfield  {number} {  }\bibfield  {author} {\bibinfo {author} {\bibfnamefont
  {K.-H.}\ \bibnamefont {Chou}}, \bibinfo {author} {\bibfnamefont {X.-J.}\
  \bibnamefont {Yu}},\ and\ \bibinfo {author} {\bibfnamefont {P.-Y.}\
  \bibnamefont {Chang}},\ }\bibfield  {title} {\bibinfo {title} {{PT
  symmetry-enriched non-unitary criticality}},\ }\Eprint
  {https://arxiv.org/abs/2509.09587} {arXiv:2509.09587} \BibitemShut {NoStop}%
\bibitem [{\citenamefont {Kaplan}\ \emph {et~al.}(2009)\citenamefont {Kaplan},
  \citenamefont {Lee}, \citenamefont {Son},\ and\ \citenamefont
  {Stephanov}}]{Kaplan-09}%
  \BibitemOpen
  \bibfield  {author} {\bibinfo {author} {\bibfnamefont {D.~B.}\ \bibnamefont
  {Kaplan}}, \bibinfo {author} {\bibfnamefont {J.-W.}\ \bibnamefont {Lee}},
  \bibinfo {author} {\bibfnamefont {D.~T.}\ \bibnamefont {Son}},\ and\ \bibinfo
  {author} {\bibfnamefont {M.~A.}\ \bibnamefont {Stephanov}},\ }\bibfield
  {title} {\bibinfo {title} {Conformality lost},\ }\href
  {https://doi.org/10.1103/PhysRevD.80.125005} {\bibfield  {journal} {\bibinfo
  {journal} {Phys. Rev. D}\ }\textbf {\bibinfo {volume} {80}},\ \bibinfo
  {pages} {125005} (\bibinfo {year} {2009})}\BibitemShut {NoStop}%
\bibitem [{\citenamefont {Wang}\ \emph {et~al.}(2017)\citenamefont {Wang},
  \citenamefont {Nahum}, \citenamefont {Metlitski}, \citenamefont {Xu},\ and\
  \citenamefont {Senthil}}]{Wang-17}%
  \BibitemOpen
  \bibfield  {author} {\bibinfo {author} {\bibfnamefont {C.}~\bibnamefont
  {Wang}}, \bibinfo {author} {\bibfnamefont {A.}~\bibnamefont {Nahum}},
  \bibinfo {author} {\bibfnamefont {M.~A.}\ \bibnamefont {Metlitski}}, \bibinfo
  {author} {\bibfnamefont {C.}~\bibnamefont {Xu}},\ and\ \bibinfo {author}
  {\bibfnamefont {T.}~\bibnamefont {Senthil}},\ }\bibfield  {title} {\bibinfo
  {title} {{Deconfined Quantum Critical Points: Symmetries and Dualities}},\
  }\href {https://doi.org/10.1103/PhysRevX.7.031051} {\bibfield  {journal}
  {\bibinfo  {journal} {Phys. Rev. X}\ }\textbf {\bibinfo {volume} {7}},\
  \bibinfo {pages} {031051} (\bibinfo {year} {2017})}\BibitemShut {NoStop}%
\bibitem [{\citenamefont {Gorbenko}\ \emph
  {et~al.}(2018{\natexlab{a}})\citenamefont {Gorbenko}, \citenamefont
  {Rychkov},\ and\ \citenamefont {Zan}}]{Gorbenko-18a}%
  \BibitemOpen
  \bibfield  {author} {\bibinfo {author} {\bibfnamefont {V.}~\bibnamefont
  {Gorbenko}}, \bibinfo {author} {\bibfnamefont {S.}~\bibnamefont {Rychkov}},\
  and\ \bibinfo {author} {\bibfnamefont {B.}~\bibnamefont {Zan}},\ }\bibfield
  {title} {\bibinfo {title} {{Walking, weak first-order transitions, and
  complex CFTs}},\ }\href {https://doi.org/10.1007/JHEP10(2018)108} {\bibfield
  {journal} {\bibinfo  {journal} {J. High Energ. Phys.}\ }\textbf {\bibinfo
  {volume} {2018}}\bibinfo  {number} { (10)},\ \bibinfo {pages}
  {108}}\BibitemShut {NoStop}%
\bibitem [{\citenamefont {Gorbenko}\ \emph
  {et~al.}(2018{\natexlab{b}})\citenamefont {Gorbenko}, \citenamefont
  {Rychkov},\ and\ \citenamefont {Zan}}]{Gorbenko-18b}%
  \BibitemOpen
\bibfield  {number} {  }\bibfield  {author} {\bibinfo {author} {\bibfnamefont
  {V.}~\bibnamefont {Gorbenko}}, \bibinfo {author} {\bibfnamefont
  {S.}~\bibnamefont {Rychkov}},\ and\ \bibinfo {author} {\bibfnamefont
  {B.}~\bibnamefont {Zan}},\ }\bibfield  {title} {\bibinfo {title} {{Walking,
  Weak first-order transitions, and Complex CFTs II. Two-dimensional Potts
  model at $Q>4$}},\ }\href {https://doi.org/10.21468/SciPostPhys.5.5.050}
  {\bibfield  {journal} {\bibinfo  {journal} {SciPost Phys.}\ }\textbf
  {\bibinfo {volume} {5}},\ \bibinfo {pages} {050} (\bibinfo {year}
  {2018}{\natexlab{b}})}\BibitemShut {NoStop}%
\bibitem [{\citenamefont {Benini}\ \emph {et~al.}(2020)\citenamefont {Benini},
  \citenamefont {Iossa},\ and\ \citenamefont {Serone}}]{Benini-20}%
  \BibitemOpen
  \bibfield  {author} {\bibinfo {author} {\bibfnamefont {F.}~\bibnamefont
  {Benini}}, \bibinfo {author} {\bibfnamefont {C.}~\bibnamefont {Iossa}},\ and\
  \bibinfo {author} {\bibfnamefont {M.}~\bibnamefont {Serone}},\ }\bibfield
  {title} {\bibinfo {title} {{Conformality Loss, Walking, and 4D Complex
  Conformal Field Theories at Weak Coupling}},\ }\href
  {https://doi.org/10.1103/PhysRevLett.124.051602} {\bibfield  {journal}
  {\bibinfo  {journal} {Phys. Rev. Lett.}\ }\textbf {\bibinfo {volume} {124}},\
  \bibinfo {pages} {051602} (\bibinfo {year} {2020})}\BibitemShut {NoStop}%
\bibitem [{\citenamefont {Faedo}\ \emph {et~al.}(2020)\citenamefont {Faedo},
  \citenamefont {Hoyos}, \citenamefont {Mateos},\ and\ \citenamefont
  {Subils}}]{Faedo-20}%
  \BibitemOpen
  \bibfield  {author} {\bibinfo {author} {\bibfnamefont {A.~F.}\ \bibnamefont
  {Faedo}}, \bibinfo {author} {\bibfnamefont {C.}~\bibnamefont {Hoyos}},
  \bibinfo {author} {\bibfnamefont {D.}~\bibnamefont {Mateos}},\ and\ \bibinfo
  {author} {\bibfnamefont {J.~G.}\ \bibnamefont {Subils}},\ }\bibfield  {title}
  {\bibinfo {title} {{Holographic Complex Conformal Field Theories}},\ }\href
  {https://doi.org/10.1103/PhysRevLett.124.161601} {\bibfield  {journal}
  {\bibinfo  {journal} {Phys. Rev. Lett.}\ }\textbf {\bibinfo {volume} {124}},\
  \bibinfo {pages} {161601} (\bibinfo {year} {2020})}\BibitemShut {NoStop}%
\bibitem [{\citenamefont {Giombi}\ \emph {et~al.}(2020)\citenamefont {Giombi},
  \citenamefont {Huang}, \citenamefont {Klebanov}, \citenamefont {Pufu},\ and\
  \citenamefont {Tarnopolsky}}]{Giombi-20}%
  \BibitemOpen
  \bibfield  {author} {\bibinfo {author} {\bibfnamefont {S.}~\bibnamefont
  {Giombi}}, \bibinfo {author} {\bibfnamefont {R.}~\bibnamefont {Huang}},
  \bibinfo {author} {\bibfnamefont {I.~R.}\ \bibnamefont {Klebanov}}, \bibinfo
  {author} {\bibfnamefont {S.~S.}\ \bibnamefont {Pufu}},\ and\ \bibinfo
  {author} {\bibfnamefont {G.}~\bibnamefont {Tarnopolsky}},\ }\bibfield
  {title} {\bibinfo {title} {{$O(N)$ model in $4 < d < 6$: Instantons and
  complex CFTs}},\ }\href {https://doi.org/10.1103/PhysRevD.101.045013}
  {\bibfield  {journal} {\bibinfo  {journal} {Phys. Rev. D}\ }\textbf {\bibinfo
  {volume} {101}},\ \bibinfo {pages} {045013} (\bibinfo {year}
  {2020})}\BibitemShut {NoStop}%
\bibitem [{\citenamefont {Gorbenko}\ and\ \citenamefont
  {Zan}(2020)}]{Gorbenko-20}%
  \BibitemOpen
  \bibfield  {author} {\bibinfo {author} {\bibfnamefont {V.}~\bibnamefont
  {Gorbenko}}\ and\ \bibinfo {author} {\bibfnamefont {B.}~\bibnamefont {Zan}},\
  }\bibfield  {title} {\bibinfo {title} {{Two-dimensional $O(n)$ models and
  logarithmic CFTs}},\ }\href {https://doi.org/10.1007/JHEP10(2020)099}
  {\bibfield  {journal} {\bibinfo  {journal} {J. High Energ. Phys.}\ }\textbf
  {\bibinfo {volume} {2020}}\bibinfo  {number} { (10)},\ \bibinfo {pages}
  {99}}\BibitemShut {NoStop}%
\bibitem [{\citenamefont {Nahum}(2022)}]{Nahum-22}%
  \BibitemOpen
\bibfield  {number} {  }\bibfield  {author} {\bibinfo {author} {\bibfnamefont
  {A.}~\bibnamefont {Nahum}},\ }\bibfield  {title} {\bibinfo {title} {{Fixed
  point annihilation for a spin in a fluctuating field}},\ }\href
  {https://doi.org/10.1103/PhysRevB.106.L081109} {\bibfield  {journal}
  {\bibinfo  {journal} {Phys. Rev. B}\ }\textbf {\bibinfo {volume} {106}},\
  \bibinfo {pages} {L081109} (\bibinfo {year} {2022})}\BibitemShut {NoStop}%
\bibitem [{\citenamefont {Han}\ \emph {et~al.}(2023)\citenamefont {Han},
  \citenamefont {Schultz},\ and\ \citenamefont {Kim}}]{Han-23}%
  \BibitemOpen
  \bibfield  {author} {\bibinfo {author} {\bibfnamefont {S.}~\bibnamefont
  {Han}}, \bibinfo {author} {\bibfnamefont {D.~J.}\ \bibnamefont {Schultz}},\
  and\ \bibinfo {author} {\bibfnamefont {Y.~B.}\ \bibnamefont {Kim}},\
  }\bibfield  {title} {\bibinfo {title} {{Complex fixed points of the
  non-Hermitian Kondo model in a Luttinger liquid}},\ }\href
  {https://doi.org/10.1103/PhysRevB.107.235153} {\bibfield  {journal} {\bibinfo
   {journal} {Phys. Rev. B}\ }\textbf {\bibinfo {volume} {107}},\ \bibinfo
  {pages} {235153} (\bibinfo {year} {2023})}\BibitemShut {NoStop}%
\bibitem [{\citenamefont {Haldar}\ \emph {et~al.}(2023)\citenamefont {Haldar},
  \citenamefont {Tavakol}, \citenamefont {Ma},\ and\ \citenamefont
  {Scaffidi}}]{Haldar-23}%
  \BibitemOpen
  \bibfield  {author} {\bibinfo {author} {\bibfnamefont {A.}~\bibnamefont
  {Haldar}}, \bibinfo {author} {\bibfnamefont {O.}~\bibnamefont {Tavakol}},
  \bibinfo {author} {\bibfnamefont {H.}~\bibnamefont {Ma}},\ and\ \bibinfo
  {author} {\bibfnamefont {T.}~\bibnamefont {Scaffidi}},\ }\bibfield  {title}
  {\bibinfo {title} {{Hidden Critical Points in the Two-Dimensional
  $\mathrm{O}(n > 2)$ Model: Exact Numerical Study of a Complex Conformal Field
  Theory}},\ }\href {https://doi.org/10.1103/PhysRevLett.131.131601} {\bibfield
   {journal} {\bibinfo  {journal} {Phys. Rev. Lett.}\ }\textbf {\bibinfo
  {volume} {131}},\ \bibinfo {pages} {131601} (\bibinfo {year}
  {2023})}\BibitemShut {NoStop}%
\bibitem [{\citenamefont {Yang}\ and\ \citenamefont {Scaffidi}()}]{Yang-26}%
  \BibitemOpen
  \bibfield  {author} {\bibinfo {author} {\bibfnamefont {C.}~\bibnamefont
  {Yang}}\ and\ \bibinfo {author} {\bibfnamefont {T.}~\bibnamefont
  {Scaffidi}},\ }\bibfield  {title} {\bibinfo {title} {{Asymptotic freedom,
  lost: Complex conformal field theory in the two-dimensional $O ( N>2 )$
  nonlinear sigma model and its realization in the spin-1 Heisenberg chain}},\
  }\Eprint {https://arxiv.org/abs/2601.02459} {arXiv:2601.02459} \BibitemShut
  {NoStop}%
\bibitem [{\citenamefont {Fisher}(1965)}]{Fisher-65}%
  \BibitemOpen
  \bibfield  {author} {\bibinfo {author} {\bibfnamefont {M.~E.}\ \bibnamefont
  {Fisher}},\ }\bibfield  {title} {\bibinfo {title} {{The Nature of Critical
  Points}},\ }in\ \href@noop {} {\emph {\bibinfo {booktitle} {Lectures in
  Theoretical Physics, Vol. {VII C}: Statistical Physics, Weak Interactions,
  Field Theory}}},\ \bibinfo {editor} {edited by\ \bibinfo {editor}
  {\bibfnamefont {W.~E.}\ \bibnamefont {Brittin}}}\ (\bibinfo  {publisher}
  {University of Colorado Press},\ \bibinfo {address} {Boulder},\ \bibinfo
  {year} {1965})\ pp.\ \bibinfo {pages} {1--159}\BibitemShut {NoStop}%
\bibitem [{\citenamefont {Ma}\ and\ \citenamefont {He}(2019)}]{Ma-19}%
  \BibitemOpen
  \bibfield  {author} {\bibinfo {author} {\bibfnamefont {H.}~\bibnamefont
  {Ma}}\ and\ \bibinfo {author} {\bibfnamefont {Y.-C.}\ \bibnamefont {He}},\
  }\bibfield  {title} {\bibinfo {title} {{Shadow of complex fixed point:
  Approximate conformality of $Q > 4$ Potts model}},\ }\href
  {https://doi.org/10.1103/PhysRevB.99.195130} {\bibfield  {journal} {\bibinfo
  {journal} {Phys. Rev. B}\ }\textbf {\bibinfo {volume} {99}},\ \bibinfo
  {pages} {195130} (\bibinfo {year} {2019})}\BibitemShut {NoStop}%
\bibitem [{\citenamefont {Jacobsen}\ and\ \citenamefont
  {Wiese}(2024)}]{Jacobsen-24}%
  \BibitemOpen
  \bibfield  {author} {\bibinfo {author} {\bibfnamefont {J.~L.}\ \bibnamefont
  {Jacobsen}}\ and\ \bibinfo {author} {\bibfnamefont {K.~J.}\ \bibnamefont
  {Wiese}},\ }\bibfield  {title} {\bibinfo {title} {{Lattice Realization of
  Complex Conformal Field Theories: Two-Dimensional Potts Model with $Q > 4$
  States}},\ }\href {https://doi.org/10.1103/PhysRevLett.133.077101} {\bibfield
   {journal} {\bibinfo  {journal} {Phys. Rev. Lett.}\ }\textbf {\bibinfo
  {volume} {133}},\ \bibinfo {pages} {077101} (\bibinfo {year}
  {2024})}\BibitemShut {NoStop}%
\bibitem [{\citenamefont {Tang}\ \emph {et~al.}(2024)\citenamefont {Tang},
  \citenamefont {Ma}, \citenamefont {Tang}, \citenamefont {He},\ and\
  \citenamefont {Zhu}}]{Tang-24}%
  \BibitemOpen
  \bibfield  {author} {\bibinfo {author} {\bibfnamefont {Y.}~\bibnamefont
  {Tang}}, \bibinfo {author} {\bibfnamefont {H.}~\bibnamefont {Ma}}, \bibinfo
  {author} {\bibfnamefont {Q.}~\bibnamefont {Tang}}, \bibinfo {author}
  {\bibfnamefont {Y.-C.}\ \bibnamefont {He}},\ and\ \bibinfo {author}
  {\bibfnamefont {W.}~\bibnamefont {Zhu}},\ }\bibfield  {title} {\bibinfo
  {title} {{Reclaiming the Lost Conformality in a Non-Hermitian Quantum 5-State
  Potts Model}},\ }\href {https://doi.org/10.1103/PhysRevLett.133.076504}
  {\bibfield  {journal} {\bibinfo  {journal} {Phys. Rev. Lett.}\ }\textbf
  {\bibinfo {volume} {133}},\ \bibinfo {pages} {076504} (\bibinfo {year}
  {2024})}\BibitemShut {NoStop}%
\bibitem [{\citenamefont {Shimizu}\ and\ \citenamefont
  {Kawabata}(2025)}]{Shimizu-25}%
  \BibitemOpen
  \bibfield  {author} {\bibinfo {author} {\bibfnamefont {H.}~\bibnamefont
  {Shimizu}}\ and\ \bibinfo {author} {\bibfnamefont {K.}~\bibnamefont
  {Kawabata}},\ }\bibfield  {title} {\bibinfo {title} {{Complex entanglement
  entropy for complex conformal field theory}},\ }\href
  {https://doi.org/10.1103/n578-ljd5} {\bibfield  {journal} {\bibinfo
  {journal} {Phys. Rev. B}\ }\textbf {\bibinfo {volume} {112}},\ \bibinfo
  {pages} {085112} (\bibinfo {year} {2025})}\BibitemShut {NoStop}%
\bibitem [{\citenamefont {Vander~Linden}\ \emph {et~al.}(2026)\citenamefont
  {Vander~Linden}, \citenamefont {De~Vos}, \citenamefont {Vervoort},
  \citenamefont {Verstraete},\ and\ \citenamefont {Ueda}}]{Linden-25}%
  \BibitemOpen
  \bibfield  {author} {\bibinfo {author} {\bibfnamefont {V.}~\bibnamefont
  {Vander~Linden}}, \bibinfo {author} {\bibfnamefont {B.}~\bibnamefont
  {De~Vos}}, \bibinfo {author} {\bibfnamefont {K.}~\bibnamefont {Vervoort}},
  \bibinfo {author} {\bibfnamefont {F.}~\bibnamefont {Verstraete}},\ and\
  \bibinfo {author} {\bibfnamefont {A.}~\bibnamefont {Ueda}},\ }\bibfield
  {title} {\bibinfo {title} {{Spiral renormalization group flow and universal
  entanglement spectrum of the non-Hermitian five-state Potts model}},\ }\href
  {https://doi.org/10.1103/l1cp-6gzr} {\bibfield  {journal} {\bibinfo
  {journal} {Phys. Rev. B}\ }\textbf {\bibinfo {volume} {113}},\ \bibinfo
  {pages} {205106} (\bibinfo {year} {2026})}\BibitemShut {NoStop}%
\bibitem [{\citenamefont {Tang}\ \emph {et~al.}(2025)\citenamefont {Tang},
  \citenamefont {Liu}, \citenamefont {Tang},\ and\ \citenamefont
  {Zhu}}]{Tang-25}%
  \BibitemOpen
  \bibfield  {author} {\bibinfo {author} {\bibfnamefont {Y.}~\bibnamefont
  {Tang}}, \bibinfo {author} {\bibfnamefont {Q.}~\bibnamefont {Liu}}, \bibinfo
  {author} {\bibfnamefont {Q.}~\bibnamefont {Tang}},\ and\ \bibinfo {author}
  {\bibfnamefont {W.}~\bibnamefont {Zhu}},\ }\bibfield  {title} {\bibinfo
  {title} {{Boundary criticality of complex conformal field theory: A case
  study in the non-Hermitian 5-state Potts model}},\ }\href
  {https://doi.org/10.21468/SciPostPhys.19.6.164} {\bibfield  {journal}
  {\bibinfo  {journal} {SciPost Phys.}\ }\textbf {\bibinfo {volume} {19}},\
  \bibinfo {pages} {164} (\bibinfo {year} {2025})}\BibitemShut {NoStop}%
\bibitem [{\citenamefont {Hikami}(1981)}]{Hikami-81}%
  \BibitemOpen
  \bibfield  {author} {\bibinfo {author} {\bibfnamefont {S.}~\bibnamefont
  {Hikami}},\ }\bibfield  {title} {\bibinfo {title} {{Three-loop
  $\beta$-functions of non-linear $\sigma$ models on symmetric spaces}},\
  }\href {https://doi.org/10.1016/0370-2693(81)90989-8} {\bibfield  {journal}
  {\bibinfo  {journal} {Phys. Lett. B}\ }\textbf {\bibinfo {volume} {98}},\
  \bibinfo {pages} {208} (\bibinfo {year} {1981})}\BibitemShut {NoStop}%
\bibitem [{\citenamefont {Wegner}(1989)}]{Wegner-89}%
  \BibitemOpen
  \bibfield  {author} {\bibinfo {author} {\bibfnamefont {F.}~\bibnamefont
  {Wegner}},\ }\bibfield  {title} {\bibinfo {title} {{Four-loop-order
  $\beta$-function of nonlinear $\sigma$-models in symmetric spaces}},\ }\href
  {https://doi.org/10.1016/0550-3213(89)90063-1} {\bibfield  {journal}
  {\bibinfo  {journal} {Nucl. Phys. B}\ }\textbf {\bibinfo {volume} {316}},\
  \bibinfo {pages} {663} (\bibinfo {year} {1989})}\BibitemShut {NoStop}%
\bibitem [{\citenamefont {Strogatz}(2024)}]{Strogatz-textbook}%
  \BibitemOpen
  \bibfield  {author} {\bibinfo {author} {\bibfnamefont {S.~H.}\ \bibnamefont
  {Strogatz}},\ }\href {https://doi.org/https://doi.org/10.1201/9780429398490}
  {\emph {\bibinfo {title} {{Nonlinear Dynamics and Chaos}}}}\ (\bibinfo
  {publisher} {Chapman and Hall/CRC, New York},\ \bibinfo {year}
  {2024})\BibitemShut {NoStop}%
\bibitem [{\citenamefont {Ueoka}\ and\ \citenamefont
  {Slevin}(2014)}]{Ueoka-Slevin-14}%
  \BibitemOpen
  \bibfield  {author} {\bibinfo {author} {\bibfnamefont {Y.}~\bibnamefont
  {Ueoka}}\ and\ \bibinfo {author} {\bibfnamefont {K.}~\bibnamefont {Slevin}},\
  }\bibfield  {title} {\bibinfo {title} {{Dimensional Dependence of Critical
  Exponent of the Anderson Transition in the Orthogonal Universality Class}},\
  }\href {https://doi.org/10.7566/JPSJ.83.084711} {\bibfield  {journal}
  {\bibinfo  {journal} {J. Phys. Soc. Jpn.}\ }\textbf {\bibinfo {volume}
  {83}},\ \bibinfo {pages} {084711} (\bibinfo {year} {2014})}\BibitemShut
  {NoStop}%
\bibitem [{\citenamefont {Ueoka}\ and\ \citenamefont
  {Slevin}(2017)}]{Ueoka-Slevin-17}%
  \BibitemOpen
  \bibfield  {author} {\bibinfo {author} {\bibfnamefont {Y.}~\bibnamefont
  {Ueoka}}\ and\ \bibinfo {author} {\bibfnamefont {K.}~\bibnamefont {Slevin}},\
  }\bibfield  {title} {\bibinfo {title} {{Borel–Padé Re-summation of the
  $\beta$-functions Describing Anderson Localisation in the Wigner-Dyson
  Symmetry Classes}},\ }\href {https://doi.org/10.7566/JPSJ.86.094707}
  {\bibfield  {journal} {\bibinfo  {journal} {J. Phys. Soc. Jpn.}\ }\textbf
  {\bibinfo {volume} {86}},\ \bibinfo {pages} {094707} (\bibinfo {year}
  {2017})}\BibitemShut {NoStop}%
\bibitem [{\citenamefont {Wang}\ \emph {et~al.}(2023)\citenamefont {Wang},
  \citenamefont {Pan}, \citenamefont {Slevin},\ and\ \citenamefont
  {Ohtsuki}}]{TongWang-23}%
  \BibitemOpen
  \bibfield  {author} {\bibinfo {author} {\bibfnamefont {T.}~\bibnamefont
  {Wang}}, \bibinfo {author} {\bibfnamefont {Z.}~\bibnamefont {Pan}}, \bibinfo
  {author} {\bibfnamefont {K.}~\bibnamefont {Slevin}},\ and\ \bibinfo {author}
  {\bibfnamefont {T.}~\bibnamefont {Ohtsuki}},\ }\bibfield  {title} {\bibinfo
  {title} {{Critical behavior of the Anderson transition in higher-dimensional
  Bogoliubov--de Gennes symmetry classes}},\ }\href
  {https://doi.org/10.1103/PhysRevB.108.144208} {\bibfield  {journal} {\bibinfo
   {journal} {Phys. Rev. B}\ }\textbf {\bibinfo {volume} {108}},\ \bibinfo
  {pages} {144208} (\bibinfo {year} {2023})}\BibitemShut {NoStop}%
\bibitem [{\citenamefont {Meurice}\ and\ \citenamefont
  {Zou}(2011)}]{Meurice-11}%
  \BibitemOpen
  \bibfield  {author} {\bibinfo {author} {\bibfnamefont {Y.}~\bibnamefont
  {Meurice}}\ and\ \bibinfo {author} {\bibfnamefont {H.}~\bibnamefont {Zou}},\
  }\bibfield  {title} {\bibinfo {title} {{Complex renormalization group flows
  for 2D nonlinear $O(N)$ sigma models}},\ }\href
  {https://doi.org/10.1103/PhysRevD.83.056009} {\bibfield  {journal} {\bibinfo
  {journal} {Phys. Rev. D}\ }\textbf {\bibinfo {volume} {83}},\ \bibinfo
  {pages} {056009} (\bibinfo {year} {2011})}\BibitemShut {NoStop}%
\bibitem [{\citenamefont {Louren\ifmmode~\mbox{\c{c}}\else \c{c}\fi{}o}\ \emph
  {et~al.}(2018)\citenamefont {Louren\ifmmode~\mbox{\c{c}}\else \c{c}\fi{}o},
  \citenamefont {Eneias},\ and\ \citenamefont {Pereira}}]{Lourenco-18}%
  \BibitemOpen
  \bibfield  {author} {\bibinfo {author} {\bibfnamefont {J.~A.~S.}\
  \bibnamefont {Louren\ifmmode~\mbox{\c{c}}\else \c{c}\fi{}o}}, \bibinfo
  {author} {\bibfnamefont {R.~L.}\ \bibnamefont {Eneias}},\ and\ \bibinfo
  {author} {\bibfnamefont {R.~G.}\ \bibnamefont {Pereira}},\ }\bibfield
  {title} {\bibinfo {title} {{Kondo effect in a $\mathcal{PT}$-symmetric
  non-Hermitian Hamiltonian}},\ }\href
  {https://doi.org/10.1103/PhysRevB.98.085126} {\bibfield  {journal} {\bibinfo
  {journal} {Phys. Rev. B}\ }\textbf {\bibinfo {volume} {98}},\ \bibinfo
  {pages} {085126} (\bibinfo {year} {2018})}\BibitemShut {NoStop}%
\bibitem [{\citenamefont {Nakagawa}\ \emph {et~al.}(2018)\citenamefont
  {Nakagawa}, \citenamefont {Kawakami},\ and\ \citenamefont
  {Ueda}}]{Nakagawa-18}%
  \BibitemOpen
  \bibfield  {author} {\bibinfo {author} {\bibfnamefont {M.}~\bibnamefont
  {Nakagawa}}, \bibinfo {author} {\bibfnamefont {N.}~\bibnamefont {Kawakami}},\
  and\ \bibinfo {author} {\bibfnamefont {M.}~\bibnamefont {Ueda}},\ }\bibfield
  {title} {\bibinfo {title} {{Non-Hermitian Kondo Effect in Ultracold
  Alkaline-Earth Atoms}},\ }\href
  {https://doi.org/10.1103/PhysRevLett.121.203001} {\bibfield  {journal}
  {\bibinfo  {journal} {Phys. Rev. Lett.}\ }\textbf {\bibinfo {volume} {121}},\
  \bibinfo {pages} {203001} (\bibinfo {year} {2018})}\BibitemShut {NoStop}%
\bibitem [{\citenamefont {Gaiotto}\ \emph {et~al.}(2021)\citenamefont
  {Gaiotto}, \citenamefont {Lee},\ and\ \citenamefont {Wu}}]{Gaiotto-21}%
  \BibitemOpen
  \bibfield  {author} {\bibinfo {author} {\bibfnamefont {D.}~\bibnamefont
  {Gaiotto}}, \bibinfo {author} {\bibfnamefont {J.~H.}\ \bibnamefont {Lee}},\
  and\ \bibinfo {author} {\bibfnamefont {J.}~\bibnamefont {Wu}},\ }\bibfield
  {title} {\bibinfo {title} {{Integrable Kondo problems}},\ }\href
  {https://doi.org/https://doi.org/10.1007/JHEP04(2021)268} {\bibfield
  {journal} {\bibinfo  {journal} {J. High Energy Phys.}\ }\textbf {\bibinfo
  {volume} {2021}}\bibinfo  {number} { (4)},\ \bibinfo {pages}
  {268}}\BibitemShut {NoStop}%
\bibitem [{\citenamefont {Nozières}\ and\ \citenamefont
  {Blandin}(1980)}]{Nozieres-80}%
  \BibitemOpen
\bibfield  {number} {  }\bibfield  {author} {\bibinfo {author} {\bibfnamefont
  {P.}~\bibnamefont {Nozières}}\ and\ \bibinfo {author} {\bibfnamefont
  {A.}~\bibnamefont {Blandin}},\ }\bibfield  {title} {\bibinfo {title} {{Kondo
  effect in real metals}},\ }\href
  {https://doi.org/https://doi.org/10.1051/jphys:01980004103019300} {\bibfield
  {journal} {\bibinfo  {journal} {J. Phys. France}\ }\textbf {\bibinfo {volume}
  {41}},\ \bibinfo {pages} {193} (\bibinfo {year} {1980})}\BibitemShut
  {NoStop}%
\bibitem [{\citenamefont {Kawabata}\ \emph {et~al.}(2019)\citenamefont
  {Kawabata}, \citenamefont {Shiozaki}, \citenamefont {Ueda},\ and\
  \citenamefont {Sato}}]{KSUS-19}%
  \BibitemOpen
  \bibfield  {author} {\bibinfo {author} {\bibfnamefont {K.}~\bibnamefont
  {Kawabata}}, \bibinfo {author} {\bibfnamefont {K.}~\bibnamefont {Shiozaki}},
  \bibinfo {author} {\bibfnamefont {M.}~\bibnamefont {Ueda}},\ and\ \bibinfo
  {author} {\bibfnamefont {M.}~\bibnamefont {Sato}},\ }\bibfield  {title}
  {\bibinfo {title} {{Symmetry and Topology in Non-Hermitian Physics}},\ }\href
  {https://doi.org/10.1103/PhysRevX.9.041015} {\bibfield  {journal} {\bibinfo
  {journal} {Phys. Rev. X}\ }\textbf {\bibinfo {volume} {9}},\ \bibinfo {pages}
  {041015} (\bibinfo {year} {2019})}\BibitemShut {NoStop}%
\bibitem [{\citenamefont {Chen}\ \emph {et~al.}(2025)\citenamefont {Chen},
  \citenamefont {Kawabata}, \citenamefont {Kulkarni},\ and\ \citenamefont
  {Ryu}}]{Chen-25}%
  \BibitemOpen
  \bibfield  {author} {\bibinfo {author} {\bibfnamefont {Z.}~\bibnamefont
  {Chen}}, \bibinfo {author} {\bibfnamefont {K.}~\bibnamefont {Kawabata}},
  \bibinfo {author} {\bibfnamefont {A.}~\bibnamefont {Kulkarni}},\ and\
  \bibinfo {author} {\bibfnamefont {S.}~\bibnamefont {Ryu}},\ }\bibfield
  {title} {\bibinfo {title} {{Field theory of non-Hermitian disordered
  systems}},\ }\href {https://doi.org/10.1103/PhysRevB.111.054203} {\bibfield
  {journal} {\bibinfo  {journal} {Phys. Rev. B}\ }\textbf {\bibinfo {volume}
  {111}},\ \bibinfo {pages} {054203} (\bibinfo {year} {2025})}\BibitemShut
  {NoStop}%
\bibitem [{\citenamefont {Berezinskii}(1971)}]{Berezinskii-71}%
  \BibitemOpen
  \bibfield  {author} {\bibinfo {author} {\bibfnamefont {V.~L.}\ \bibnamefont
  {Berezinskii}},\ }\bibfield  {title} {\bibinfo {title} {{Destruction of
  long-range order in one-dimensional and two-dimensional systems having a
  continuous symmetry group. I. Classical systems}},\ }\href@noop {} {\bibfield
   {journal} {\bibinfo  {journal} {J. Exp. Theor. Phys.}\ }\textbf {\bibinfo
  {volume} {32}},\ \bibinfo {pages} {493} (\bibinfo {year} {1971})}\BibitemShut
  {NoStop}%
\bibitem [{\citenamefont {Berezinskii}(1972)}]{Berezinskii-72}%
  \BibitemOpen
  \bibfield  {author} {\bibinfo {author} {\bibfnamefont {V.~L.}\ \bibnamefont
  {Berezinskii}},\ }\bibfield  {title} {\bibinfo {title} {{Destruction of
  long-range order in one-dimensional and two-dimensional systems having a
  continuous symmetry group. II. Quantum systems}},\ }\href@noop {} {\bibfield
  {journal} {\bibinfo  {journal} {J. Exp. Theor. Phys.}\ }\textbf {\bibinfo
  {volume} {34}},\ \bibinfo {pages} {610} (\bibinfo {year} {1972})}\BibitemShut
  {NoStop}%
\bibitem [{\citenamefont {Kosterlitz}\ and\ \citenamefont
  {Thouless}(1973)}]{Kosterlitz-Thouless-73}%
  \BibitemOpen
  \bibfield  {author} {\bibinfo {author} {\bibfnamefont {J.~M.}\ \bibnamefont
  {Kosterlitz}}\ and\ \bibinfo {author} {\bibfnamefont {D.~J.}\ \bibnamefont
  {Thouless}},\ }\bibfield  {title} {\bibinfo {title} {{Ordering, metastability
  and phase transitions in two-dimensional systems}},\ }\href
  {https://doi.org/10.1088/0022-3719/6/7/010} {\bibfield  {journal} {\bibinfo
  {journal} {J. Phys. C}\ }\textbf {\bibinfo {volume} {6}},\ \bibinfo {pages}
  {1181} (\bibinfo {year} {1973})}\BibitemShut {NoStop}%
\bibitem [{\citenamefont {Kosterlitz}(1974)}]{Kosterlitz-74}%
  \BibitemOpen
  \bibfield  {author} {\bibinfo {author} {\bibfnamefont {J.~M.}\ \bibnamefont
  {Kosterlitz}},\ }\bibfield  {title} {\bibinfo {title} {{The critical
  properties of the two-dimensional XY model}},\ }\href
  {https://doi.org/10.1088/0022-3719/7/6/005} {\bibfield  {journal} {\bibinfo
  {journal} {J. Phys. C}\ }\textbf {\bibinfo {volume} {7}},\ \bibinfo {pages}
  {1046} (\bibinfo {year} {1974})}\BibitemShut {NoStop}%
\bibitem [{\citenamefont {Plenio}\ and\ \citenamefont
  {Knight}(1998)}]{Plenio-review}%
  \BibitemOpen
  \bibfield  {author} {\bibinfo {author} {\bibfnamefont {M.~B.}\ \bibnamefont
  {Plenio}}\ and\ \bibinfo {author} {\bibfnamefont {P.~L.}\ \bibnamefont
  {Knight}},\ }\bibfield  {title} {\bibinfo {title} {{The quantum-jump approach
  to dissipative dynamics in quantum optics}},\ }\href
  {https://doi.org/10.1103/RevModPhys.70.101} {\bibfield  {journal} {\bibinfo
  {journal} {Rev. Mod. Phys.}\ }\textbf {\bibinfo {volume} {70}},\ \bibinfo
  {pages} {101} (\bibinfo {year} {1998})}\BibitemShut {NoStop}%
\bibitem [{\citenamefont {Daley}(2014)}]{Daley-review}%
  \BibitemOpen
  \bibfield  {author} {\bibinfo {author} {\bibfnamefont {A.~J.}\ \bibnamefont
  {Daley}},\ }\bibfield  {title} {\bibinfo {title} {{Quantum trajectories and
  open many-body quantum systems}},\ }\href
  {https://doi.org/10.1080/00018732.2014.933502} {\bibfield  {journal}
  {\bibinfo  {journal} {Adv. Phys.}\ }\textbf {\bibinfo {volume} {63}},\
  \bibinfo {pages} {77} (\bibinfo {year} {2014})}\BibitemShut {NoStop}%
\end{thebibliography}

%

\end{document}